\documentclass[%
 reprint,
 longbibliography,
 superscriptaddress,
 amsmath,
 amssymb,
 aps,
 prx,
 floatfix,
]{revtex4-2}
\usepackage[utf8]{inputenc}
\pdfoutput=1

\usepackage{graphicx}
\usepackage{dcolumn}
\usepackage{bm}
\usepackage[dvipsnames]{xcolor}
\usepackage[colorlinks=true,linkcolor=teal,citecolor=Maroon,urlcolor=Maroon]{hyperref}
\usepackage{physics}
\usepackage{siunitx}
\usepackage{float}
\usepackage{tikz}
\usetikzlibrary{quantikz}
\usepackage{mathbbol}
\usepackage{enumitem}
\usepackage{xfrac}
\newcommand{\qgate}[1]{\textsc{#1}}
\newcommand{\qmeas}[1]{\textsc{#1}}
\newcommand{\qcode}[1]{\texttt{#1}}
\newcommand{\avgFid}{\overline{\mathcal{F}}}
\newcommand{\entFid}{\mathcal{F}_{\mathcal{E}}}

\newcommand{\secref}[1]{Section~\hyperref[#1]{\ref*{#1}}}
\newcommand{\appref}[1]{Appendix~\hyperref[#1]{\ref*{#1}}}
\newcommand{\tabref}[1]{Table~\hyperref[#1]{\ref*{#1}}}
\newcommand{\figref}[1]{Fig.~\hyperref[#1]{\ref*{#1}}}
\newcommand{\sfigref}[2]{Fig.~\hyperref[#1]{\ref*{#1}(#2)}}

\usepackage{svg}
\newcommand{\orcid}[1]{\href{https://orcid.org/#1}{\includegraphics[width=8pt]{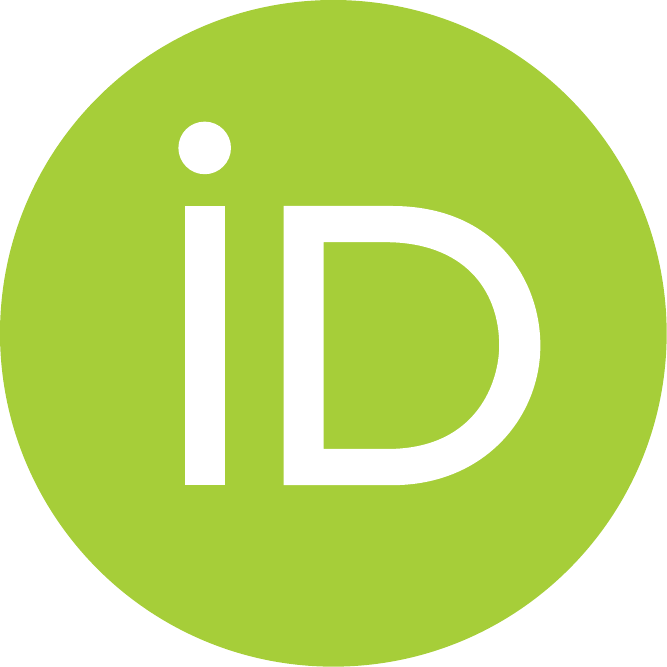}}}
\newcommand{\timo}[1]{#1}
\newcommand{\timorevised}[1]{{#1}}

\begin{document}

\title{Performance of teleportation-based error correction circuits for bosonic codes with noisy measurements}
\author{Timo Hillmann\orcid{0000-0002-1476-0647}}
\email{hillmann@chalmers.se}
\affiliation{Department of Microtechnology and Nanoscience (MC2), Chalmers University of Technology, SE-412 96 Gothenburg, Sweden}
\author{Fernando Quijandría\orcid{0000-0002-2355-0449}
}
\affiliation{Department of Microtechnology and Nanoscience (MC2), Chalmers University of Technology, SE-412 96 Gothenburg, Sweden}

\author{Arne L. Grimsmo\orcid{0000-0002-5208-0271}
}
\affiliation{Centre for Engineered Quantum Systems, School of Physics, The University of Sydney, Sydney, Australia}

\author{Giulia Ferrini\orcid{0000-0002-7130-6723}}
\affiliation{Department of Microtechnology and Nanoscience (MC2), Chalmers University of Technology, SE-412 96 Gothenburg, Sweden}

\date{\today}
\begin{abstract}
    Bosonic quantum error-correcting codes offer a viable direction towards reducing the hardware overhead required for fault-tolerant quantum information processing.
    A broad class of bosonic codes, namely rotation-symmetric codes, can be characterized by their phase-space rotation symmetry.
    However, their performance has been examined to date only within an idealistic noise model.
    Here, we further analyze the error correction capabilities of rotation-symmetric codes using a teleportation-based error correction circuit.
    To this end, we numerically compute the average gate fidelity including measurement errors into the noise model of the data qubit. Focusing on physical measurements models, we assess the performance of heterodyne and adaptive homodyne detection in comparison to the previously studied canonical phase measurement.
    This setting allows to shed light on the role of different currently available measurement schemes when decoding the encoded information.
    We find that with the currently achievable measurement efficiencies in microwave optics, bosonic rotation codes undergo a substantial decrease in their break-even potential.
    \timo{
    In addition, we perform a detailed analysis of Gottesman-Kitaev-Preskill codes using a similar error correction circuit which allows us to analyze the effect of realistic measurement models on different codes.
    In comparison to RSB codes, we find for GKP codes an even greater reduction in performance together with a vulnerability to photon number dephasing.}
    Our results show that highly efficient measurement protocols constitute a crucial building block towards error-corrected quantum information processing with bosonic continuous-variable systems.
\end{abstract}

\maketitle

\section{Introduction\label{sec:intrdocution}}

The principle of quantum error correction (QEC) is to encode the logical information redundantly in a subspace of a much larger Hilbert space.
Bosonic codes that make use of the infinite-dimensional Hilbert space of a single bosonic mode~\cite{chuang_bosonic_1997, cochrane_macroscopically_1999, gottesman_encoding_2001, ralph_quantum_2003} have emerged as a hardware efficient alternative to conventional codes utilizing multiple two-level systems~\cite{shor_scheme_1995, shor_fault-tolerant_1996, gottesman_introduction_2009, terhal_quantum_2015}.
This extended Hilbert space of a single bosonic modes enables one to tailor the encoding of the logical information
to a specific noise channel~\cite{lund_fault-tolerant_2008, mirrahimi_dynamically_2014, michael_new_2016, li_cat_2017, li_phase-engineered_2021, puri_stabilized_2019, guillaud_error_2021}, adapted to the physical system in which the bosonic code is realized.
Despite the flexibility in choosing the encoding, many bosonic codes belong to a larger class of codes which is characterized by a discrete rotation symmetry~\cite{grimsmo_quantum_2020}. 
These instances of bosonic codes are called rotation-symmetric bosonic (RSB) codes or bosonic rotation codes.  

From an experimental perspective, bosonic codes have advanced considerably in the recent decades and various code states such as cat states~\cite{cochrane_macroscopically_1999, mirrahimi_dynamically_2014, bergmann_quantum_2016}, binomial states~\cite{michael_new_2016}, and Gottesman-Kitaev-Preskill (GKP) states~\cite{gottesman_encoding_2001} have been realized in electromagnetic modes of superconducting microwave circuits~\cite{vlastakis_deterministically_2013, leghtas_confining_2015, wang_heisenberg-limited_2019, grimm_stabilization_2020} and motional modes of trapped ions~\cite{kienzler_observation_2016, fluhmann_encoding_2019, de_neeve_error_2022}.
Importantly, it has been shown experimentally that bosonic codes can reach, and even go beyond, the break-even point for error correction~\cite{ofek_extending_2016,hu_quantum_2019}, which is the point where the error corrected qubit's lifetime is as good as that of an unencoded qubit realized in the same hardware.
Experimental progress towards fault-tolerant error correction has been made recently as well~\cite{rosenblum_fault-tolerant_2018, hu_quantum_2019, ma_error-transparent_2020, ma_path-independent_2020, reinhold_error-corrected_2020}.
However, in the current literature on error correcting schemes with bosonic codes the error model typically focuses on state preparation errors~\cite{glancy_error_2006, noh_low-overhead_2022, bourassa_blueprint_2021}, while gate and measurement errors are either excluded or described by a simplified error model to make the analysis less unwieldy. Only recently, studies have begun focusing on the practical realization of qubit gates for GKP codes with finite energy~\cite{tzitrin_progress_2020, hastrup_unsuitability_2021, grimsmo_gkp_2021}.

In this work we study single-mode bosonic qubits undergoing simultaneous loss and dephasing followed by error correction. Our main goal is to shed light on the role of different currently available measurement schemes when decoding the encoded information, and we therefore model other elements, such as state preparation, gates and the ancillae used in error correction circuit to be essentially noiseless. In particular, this allows us to directly compare to the previous results in Ref.~\cite{grimsmo_quantum_2020}. Moreover, this setting may serve as a model for a quantum communication scenario where a bosonic code is used to transmit information over a noise channel, while the operations used for encoding and decoding the information can be done with high fidelity.

In particular, we consider error correction of RSB and GKP codes through a Knill-type teleportation-based error correction circuit which has seen an increased interest recently~\cite{grimsmo_quantum_2020, walshe_continuous-variable_2020, larsen_fault-tolerant_2021, noh_low-overhead_2022}.
For RSB codes we numerically investigate the error correction potential of the scheme proposed in Ref.~\cite{grimsmo_quantum_2020} by computing the average gate fidelity when the logical \qgate{X} measurements are realized by (finite-efficiency) heterodyne and adaptive homodyne measurements and compare the results to the ideal but unphysical canonical phase measurement that was considered in Ref.~\cite{grimsmo_quantum_2020}.
To this end, we assume that the noisy encoded qubit is the result of a simultaneous loss and dephasing channel before the above-mentioned error-correction circuit is applied.
Our numerical results show that RSB codes can in principle still reach the break-even point for error correction, but undergo a substantial decrease in performance for state-of-the-art measurement schemes.
\timo{
In the second part of the article, we extend our analysis to Gottesman-Kitaev-Preskill (GKP) codes using a similar error correction circuit in which homodyne detection takes the role of logical measurements in order to compare the effect of realistic measurement models on different codes.
GKP codes are theoretically expected to outperform RSB codes if the noise channel consists only of single-photon losses and error-correction is implemented via the optimal recovery map~\cite{albert_performance_2018, noh_quantum_2019}.
However, here we not only identify an even greater sensitivity to finite measurement efficiencies for GKP codes compared to RSB codes, but also a vulnerability to photon number dephasing noise.
We find that within this extended noise model the performance of GKP codes can fall behind that of RSB codes in experimentally relevant parameter regimes.
In addition, at the level of methodology, we show that for GKP codes the average gate fidelity can be exactly related to the logical success probability if the noise is characterized by Gaussian random displacement channels and no noise propagates to the output system.
This equivalence allows for a quantitative comparison between numerically exact \emph{bonafide} finite-energy GKP states with an approximate analytical model obtained from the twirling approximation.
Importantly, we find that the two approaches give the same results in this setting.
This comparison between the two approaches has been missing in the literature since the introduction of the state-twirling approximation for the description of GKP states in Ref.~\cite{menicucci_fault-tolerant_2014}.
As this approximation is the basis for current studies of fault-tolerance thresholds with topological-GKP codes~\cite{noh_fault-tolerant_2020, noh_low-overhead_2022, larsen_fault-tolerant_2021, bourassa_blueprint_2021} our study significantly corroborates these results.
}

The subsequent sections are organized as follows.
In \secref{sec:rsb} we review basic properties and notations of rotation-symmetric bosonic codes, including logical operations as well as propagation and correction of single-mode errors.
We give definitions for the sub-classes of RSB codes that we have analyzed in this paper, namely \qcode{cat} and \qcode{bin} codes, in \secref{ssec:explicit_rsb_codes}.
The setup and noise model that is studied is laid out in \secref{sec:noise_model} and the different realizations of the logical \qgate{X} measurements in terms of phase measurements are described in \secref{ssec:phase_measurements}.
Our numerical results for RSB codes are presented in \secref{sec:num_res_rsb}.
To allow for a comparison to finite-energy GKP codes we first introduce their basic properties and a slightly modified error model in \secref{sec:gkp_comparison} before presenting the results at the end of the section.
We end the paper with a discussion and an outlook in \secref{sec:discussion_conclusion}.

\section{Rotation-Symmetric Bosonic Codes\label{sec:rsb}}
A bosonic encoding has an $N$-fold rotation symmetry if its two-dimensional code space projector, 
\begin{align}
    \label{eq:code_projector_general}
    \hat{\Pi}_{\qcode{code}} = \ketbra{0}_{\qcode{code}} + \ketbra{1}_{\qcode{code}},
\end{align}
commutes with the discrete rotation operator
\begin{align}
    \label{eq:discrete_rot_op}
    \hat{R}_N = e^{i 2 \pi \hat{n} / N},
\end{align}
where $\hat{n} = \hat{a}^{\dagger} \hat{a}$ is the number operator, that is, $\hat{n} \ket{n} = n \ket{n}$ for Fock states $\ket{n}, n \in \mathbb{N}_0$.
Here, $\hat{a}$ and $\hat{a}^{\dagger}$ denote the annihilation and creation operator of a single bosonic mode, respectively.
They obey the canonical commutation relation  $\comm{\hat{a}}{\hat{a}^{\dagger}}= \mathbb{1}$.

We define an order-$N$ rotation-symmetric bosonic (RSB) code through the additional requirement that the operator
\begin{align}
    \label{eq:logical_Z_op}
    \hat{Z}_N = \hat{R}_{2 N} = e^{i \pi \hat{n} / N},
\end{align}
acts as the logical $\qgate{Z}$ gate with eigenvalues $\pm 1$ on the code subspace defined by $\hat{\Pi}_{\qcode{code}}$. (The choice of logical \qgate{Z} over logical \qgate{X} is of course arbitrary, but we do not consider codes where $\hat R_{2N}$ acts as a non-Pauli gate, such as GKP codes~\cite{grimsmo_quantum_2020}, to be RSB codes.)

\begin{figure}[t]
    \centering
    \includegraphics{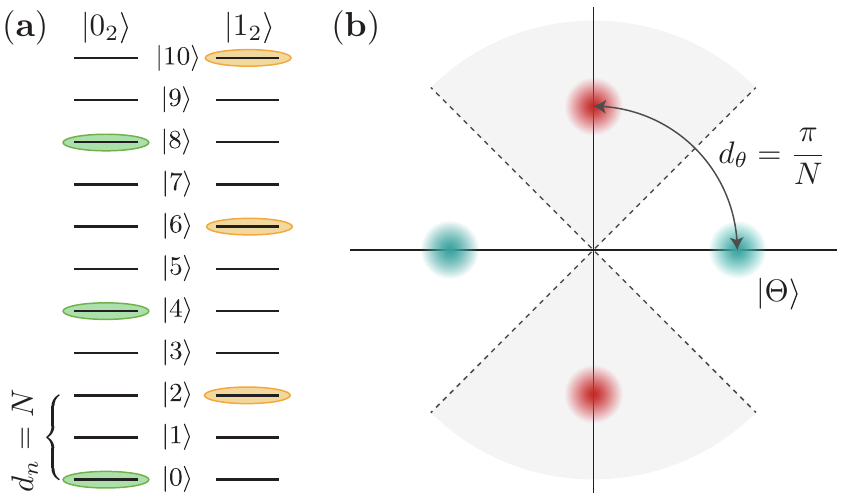} 
    \caption{Fock \timo{space} and phase-sector structure of RSB codes with order-$N=2$ rotation symmetry. (a) The computational states of order-$N$ codes have support only on each $2N$-th Fock state, with the $\ket{1_N}$ state offset by $N$ from the $\ket{0_N}$ state. This introduces a number distance $d_n = N$ between the codewords. (b) The dual basis code states $\ket{\pm_N}$ can be constructed as the superpositions of $N$ rotated primitive states  $\ket{\Theta}$. Here, we associate the blue (red) primitive states to the $\ket{+_2}$ ($\ket{-_2}$) dual-basis code word which occupies the white (gray) phase-sectors exclusively. The number distance $d_n$ of computational states $\ket{\mu_N}$ translates to a rotational distance $d_{\theta} = \pi / N$ for dual-basis states $\ket{\pm_N}$.}
    \label{fig:fock_phase_space_structure}
\end{figure}

This requirement allows for constructing the computational states $\ket{\mu_N}_{\qcode{code}} (\mu = 0, 1)$ of any order-$N$ rotation code in terms of a finite superposition of rotated normalized primitive states $(\hat{R}_N)^m \ket{\Theta}$.
That is, $\hat{Z}_N \ket{\mu_N}_{\qcode{code}} = (-1)^{\mu} \ket{\mu_N}_{\qcode{code}}$ if and only if the computational basis states can be written as~\cite{grimsmo_quantum_2020}
\begin{align}
    \label{eq:rsb_rotated_primitive}
    \ket{\mu_N}_{\qcode{code}} = \frac{1}{\sqrt{\mathcal{N}_\mu}} \sum_{m = 0}^{2 N - 1} (-1)^{\mu m} e^{i \pi m \hat{n} / N} \ket{\Theta},
\end{align}
with normalization constants $\mathcal{N}_{\mu}$.
\timo{Such a primitive state $\ket{\Theta}$ uniquely defines an order-N RSB code and we can therefore fully describe any RSB code through its primitive state and its rotation symmetry $N$.}
The constraint on the primitive state $\ket{\Theta}$ is that it must have at least partial support on the code states, that is, $\braket{2 k N}{\Theta} \neq 0$ and $\braket{(2 k^{\prime} + 1) N}{\Theta} \neq 0$ for some integers $k$ and $k^{\prime}$, respectively.
This constraint originates from the Fock \timo{ space} structure of the code states $\ket{\mu_N}_{\qcode{code}}$ that is enforced through the operator $\hat{Z}_N$ and illustrated in \sfigref{fig:fock_phase_space_structure}{a}.
Therefore, it is straight-forward to realize that the computational states can be written alternatively as~\cite{grimsmo_quantum_2020}
\begin{align}
        \label{eq:logical_rsb_states}
        \ket{\mu_N}_{\qcode{code}} = \sum_{k = 0}^{\infty} f_{(2 k + \mu) N} \ket{(2 k + \mu) N},
\end{align}
with Fock \timo{ space} coefficients $f_{k N}$.
Observe that this yields a Fock \timo{ space} distance $d_n = N$ between the logical code states $\ket{\mu_N}_{\qcode{code}}$ which is illustrated in \sfigref{fig:fock_phase_space_structure}{a}.
Furthermore, if $\lvert f_{k N} \rvert = \lvert f_{(k+1)N} \rvert$ for all $k$, we refer to the rotational code as a number-phase code.
The reason is that these codes have a vanishing phase uncertainty as we will discuss in more details in \secref{ssec:phase_measurements}.
\timo{Because (ideal) number-phase codes are unphysical, we instead refer to codes that satisfy the above relation in the limit $\overline{n}_{\qcode{code}} \to \infty$ as \emph{approximate} number-phase codes, see \secref{ssec:explicit_rsb_codes} for examples.}

In contrast to the computational states $\ket{\mu_N}_{\qcode{code}}$, the dual-basis code states, 
\begin{align}
    \label{eq:rsb_dual_basis_code}
   \ket{\pm_N}_{\qcode{code}}= (\ket{0_N}_{\qcode{code}} \pm \ket{1_N}_{\qcode{code}}) / \sqrt{2}, 
\end{align}
have support on each Fock state $\ket{k N}$ and can be represented as
\begin{align}
    \ket{\pm_N} = \frac{1}{\sqrt{2}} \sum_{k =0}^{\infty} (\pm 1)^{k} f_{k N} \ket{k N}.
\end{align}
The $\ket{+_N}_{\qcode{code}}$ ($\ket{-_N}_{\qcode{code}}$) state is constructed from primitive states $\ket{\Theta}$ that are rotated by an angle $\frac{\pi m}{N}$ with $m$ even (odd) as evident from Eq.~\eqref{eq:rsb_rotated_primitive}.
This means that the two states are separated by a ``rotational distance'' $d_{\theta} = \pi / N$ in phase space. In contrast, the computational states $\ket{\mu_N}$ have zero rotational distance and occupy the same phase sectors.
This is illustrated in \sfigref{fig:fock_phase_space_structure}{b}.
\subsection{Logical operations\label{ssec:rsb_logical_operations}}
Quantum computing schemes with number-phase codes are based on the availability of an encoded $\qgate{CZ} = \mathrm{diag}(1, 1, 1, -1)$ gate, state preparation $\mathcal{P}_{\ket{\Psi_N}}$ of encoded states $\ket{\Psi_N}$, and measurements $\mathcal{M}_{\qgate{X}}$ in the logical $\ket{\pm_N}$ basis.
The \qgate{CZ} gate between two RSB codes of order-$N$ and order-$M$ corresponds to a controlled rotation (\qgate{CROT}), that is,
\begin{align}
    \label{eq:rsb_gate_crot}
    \qgate{CZ}_{NM} \equiv \qgate{CROT}_{NM} = \exp(i \frac{\pi}{NM} \hat{n} \otimes \hat{n}).
\end{align}
With the entangling \qgate{CROT} gate any gate $\qgate{G} \qgate{H}$, with $\qgate{G} = \mathrm{diag}(\lambda_1, \lambda_2)$ diagonal in the computational basis ($\lvert \lambda_1\rvert^2 = \lvert \lambda_2 \rvert^2 = 1$) and \qgate{H} the Hadamard gate, can be executed using the following teleportation circuit
\begin{equation}
    \label{eq:gate_teleportation_circuit}
    \begin{quantikz}
    \lstick{$\ket{\Psi_N}$} & \gate[2, disable auto height]{\rotatebox{270}{CROT}} &  \meterD{\mathcal{M}_X} \rstick{$\pm$} \\
    \lstick{$\ket{G_M}$} &  & \qw  \rstick{$\qgate{G} \qgate{X}^{i} \qgate{H} \ket{\psi_M},$}
    \end{quantikz}
\end{equation}
where the ancilla system is prepared in the state $\ket{G_M} = (\lambda_1 \ket{0_M} + \lambda_2 \ket{1_M}) / \sqrt{2}$. 
Here and in the following we use the convention $i=0$ for the measurement outcome ``$+$'' and $i=1$ for the outcome ``$-$''.
To achieve universality it is sufficient to be able to prepare the states $\ket{+_M}$, $\ket{T_M} = (\ket{0}_M + e^{i \pi / 4} \ket{1_M}) / \sqrt{2}$, and $\ket{+i_M} = (\ket{0}_M + i \ket{1_M}) / \sqrt{2}$ which yield \timo{for G} the identity~\footnote{Alternatively, one can think about $\ket{+_M}$ teleporting the \qgate{H} gate by post-selecting on the $i=0$ outcome.}, the non-Clifford $\qgate{T}$-gate, and the phase gate $\qgate{S}$, respectively.

\subsection{Error correction \label{ssec:rsb_error_correction}}
Any practical implementation of a quantum computing scheme will suffer from errors.
Even though small errors, i.e., errors that do not lead to a false decoding of the logical information, are acceptable for bosonic encodings~\cite{gottesman_encoding_2001}, it is still necessary to perform error correction periodically to prevent the accumulation of small errors into such a logical error.
As was shown in Ref.~\cite{grimsmo_quantum_2020}, implementing the universal set of operations through the teleportation circuit in Eq.~\eqref{eq:gate_teleportation_circuit} already ensures that small errors will not be amplified too badly when they propagate through the circuit.
We review the method of error correction by teleportation for RSB codes in the following and describe the propagation of errors in \appref{app:rsb_error_propagation} for completeness.

\begin{figure}[b]
    \centering
    \begin{quantikz}
    \lstick{$\mathcal{N}(\hat{\rho}_N)$} &  \gate[2, disable auto height]{\rotatebox{270}{CROT}}  &\meterD{\mathcal{M}_X} \rstick{$x_1$} \\
    \lstick{$\ket{+_M}$} & & \gate[2, disable auto height]{\rotatebox{270}{CROT}} & \meterD{\mathcal{M}_X} \rstick{$x_2$} \\
    \lstick{$\ket{+_L}$} & \qw & & \qw \rstick{$\mathcal{R}^{\mathrm{Knill}} \circ \mathcal{N}(\hat{\rho}_N) $}
    \end{quantikz}
    \caption{Bosonic version of Knill-type error correction for number-phase codes. In the case of noise free ancillae the circuit implements a logical (qubit to qubit) channel, also referred to as the telecorrection circuit. The recovery map $\mathcal{R}^{\mathrm{Knill}}$ corresponds to a change of the Pauli frame and is determined by the measurement outcomes $\vec{x} = (x_1, x_2)$, see main text for details.
    Thus, a logical error on the output state occurs only if the decoding of the $\mathcal{M}_{\qgate{X}}$ measurement is faulty.}
    \label{fig:telecorrection}
\end{figure}
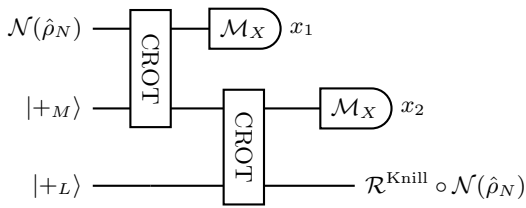

Error correction can be implemented by sequentially performing two one-bit teleportations [cf. Eq.~\eqref{eq:gate_teleportation_circuit}] with ancillae prepared in $\ket{+_M}$ and $\ket{+_L}$, respectively~\cite{grimsmo_quantum_2020}.
\footnotetext[12]{See also \appref{sec:knill_ec} for a summary of Knill-type error correction in the discrete-variable case.}
This represents a continuous-variable version of Knill-type error correction~\cite{knill_scalable_2005, knill_quantum_2005}, see also \appref{sec:knill_ec} for a summary of Knill-type error correction in the discrete-variable case.
Under the premise that the ancilla $\ket{+_M}$ does not undergo any number-shift errors \emph{before} the \qgate{CROT} gates are executed and that the second ancilla $\ket{+_L}$ is noise free, the resulting quantum channel $\mathcal{R}^{\mathrm{Knill}} \circ \mathcal{N}$ of the teleportation circuit in \figref{fig:telecorrection} is a logical (qubit to qubit) channel~\cite{grimsmo_quantum_2020}.
Here, we denoted by $\mathcal{N}$ the initial noise channel.
In other words, the resulting output qubit state $\hat{\rho}_L = \mathcal{R}^{\mathrm{Knill}} \circ \mathcal{N}(\hat{\rho}_N) $ encoded into an order-$L$ rotation code represents the same logical state as the initial logical state $\hat{\rho}_N$ that was encoded into an order-$N$ rotation code up to logical errors.
That is, the output state takes the form~\cite{grimsmo_quantum_2020}
\begin{align}
    \label{eq:rsb_knill_ec_out}
    \mathcal{R}^{\text{Knill}} \circ \mathcal{N}(\hat{\rho}_N) = \frac{1}{4} \sum_{i, j = 0}^{3} \sum_{\vec{x}} c_{ij}(\vec{x}) \qgate{P}_{i^*(\vec{x})}^{\dagger}  \qgate{P}_i \hat{\rho}_L \qgate{P}^{\dagger}_j \qgate{P}_{i^*(\vec{x})},
\end{align}
where the weights $c_{ij}(\vec{x}) = \Tr[\hat{M}_{x_1} \otimes \hat{M}_{x_2} \hat{\sigma}_{ij}]$, associated to measurement outcomes $\vec{x} = (x_1, x_2)$, depend on the realization of the $\mathcal{M}_{\qgate{X}}$ measurement described by the POVM elements $\lbrace \hat{M}_{x_{1/2}} \rbrace$ and the operators $\qgate{P}_i \in \lbrace \qgate{I}, \qgate{Z}, \qgate{X}, \qgate{X} \qgate{Z} \rbrace$ represent logical Pauli operators on the encoded output state $\hat{\rho}_L$.
The operator $\hat{\sigma}_{ij}$ represents the damaged two-mode dual-basis code states that are obtained by commuting the noise channel $\mathcal{N}$ through the \qgate{CROT} gate, that is,
\begin{align}
    \hat{\sigma}_{ij} = \mathcal{U}_{\qgate{CROT}} \circ \mathcal{N} \circ \mathcal{U}_{\qgate{CROT}}^{\dagger}(\ketbra{i}{j}),
\end{align}
where $\ket{i} = \qgate{H}\ket{a_N} \otimes \qgate{H}\ket{b_M}$, $ab$ is the binary representation of $i$, and $\mathcal{U}_{\qgate{CROT}} \boldsymbol{\cdot} = \qgate{CROT} \boldsymbol{\cdot} \qgate{CROT}^{\dagger}$.
The problem of recovering the original logical state reduces to determining the correct Pauli recovery $\qgate{P}_x$ for each application of the error-correction circuit.
As in Ref.~\cite{grimsmo_quantum_2020}, we guess the most likely Pauli correction $\qgate{P}_{i^{*}(\vec{x})}$ based on a maximum likelihood decoder 
\begin{align}
    \label{eq:rsb_decoder}
    i^{*}(\vec{x}) = \max_{i=0,\dots,3} \Tr[ \hat{M}_{x_1} \otimes \hat{M}_{x_2} \hat{\sigma}_{ii}].
\end{align}


\subsection{Examples of RSB codes \label{ssec:explicit_rsb_codes}}
We restrict our analysis to two classes of rotational number-phase codes, namely, \qcode{bin} (binomial) and \qcode{cat} codes.
Our choice of these codes is motivated by the following observations.
Firstly, the results of Ref.~\cite{grimsmo_quantum_2020} indicate that \qcode{bin} codes outperform almost all other number-phase codes.
Secondly, the increased interest in dissipatively engineered \qcode{cat} states~\cite{lescanne_exponential_2020, chamberland_building_2022, puri_stabilized_2019, guillaud_repetition_2019} motivates further analysis of those codes.
Furthermore, \qcode{cat} states with up to $100$ photons have already been prepared in experiments~\cite{vlastakis_deterministically_2013} whereas to our knowledge the largest prepared binomial state contained only a few photons.
This observation is relevant because the results of Ref.~\cite{grimsmo_quantum_2020} as well as \secref{sec:num_res_rsb} indicate performance sweet spots for relatively large 
average code photon number
$\overline{n}_{\qcode{code}} = \Trace (\hat{\Pi}_{\qcode{code}} \hat{n}) / 2$ in many cases.

\subsubsection{\qcode{cat} codes\label{sssec:cat_codes}}
\qcode{cat} codes are the earliest form of single-mode RSB codes~\cite{cochrane_macroscopically_1999}.
The computational code states $\ket{\mu_N}_{\qcode{cat}}$ are most straight-forwardly constructed from their primitive which are coherent states $\ket{\Theta}_{\qcode{cat}} = \ket{\alpha} = e^{- \lvert \alpha \rvert^2 / 2} \sum_k \frac{\alpha^k}{\sqrt{k!}} \ket{k}$, and $\alpha > 0$ without loss of generality.
The rotated superposition of the primitive yields the computational basis states,
\begin{align}
\label{eq:logical_basis_cat_codes}
\ket{\mu_N}_{\qcode{cat}} &= \frac{1}{\sqrt{\mathcal{N}_{\mu}}} \sum_{m = 0}^{2 N - 1} (-1)^{\mu m} \ket*{\alpha e^{i m \pi / N}},
\end{align}
with normalization constants $\mathcal{N}_{\mu}$.
From the Fock \timo{ space} representation of the coherent state, the coefficients $\lbrace f_{k N} \rbrace$ in Eq.~\eqref{eq:logical_rsb_states} are obtained as
\begin{align}
\label{eq:fock_grid_coeff_cat_codes}
    f_{k N} =\sqrt{ \frac{2 N^2}{\mathcal{N}_i}} e^{- \lvert \alpha \rvert^2 / 2} \frac{\alpha^{k N}}{\sqrt{(k N)!}},
\end{align}
with $\mathcal{N}_i = \mathcal{N}_0$ and $\mathcal{N}_i = \mathcal{N}_1$ for even and odd $k$, respectively.

\subsubsection{\qcode{bin} codes\label{sssec:bin_codes}}
The class of \qcode{bin} codes was first introduced in Ref.~\cite{michael_new_2016}, designed explicitly to correct against photon loss, photon gain, and photon dephasing errors up to a certain order.
Their name originates from the binomial coefficient that weights each Fock state that is used to construct the code words, that is,
\begin{align}
\label{eq:logical_basis_bin_codes}
\ket{\mu_N}_{\qcode{bin}} &= \sum_{k = 0}^{\left \lfloor{K/2} \right \rfloor - \mu} \sqrt{\frac{1}{2^{K-1}} \mqty(K \\ 2 k + \mu)} \ket{(2 k + \mu) N}, 
\end{align}
for the computational basis code states $\ket{\mu_N}_{\qcode{bin}}$.
Here the truncation parameter $K$ relates to the number of errors that are exactly correctable, see Ref.~\cite{michael_new_2016} for details.
By comparison to Eq.~\eqref{eq:logical_rsb_states} the Fock \timo{ space} coefficients $\lbrace f_{k N} \rbrace$ can be directly identified from Eq.~\eqref{eq:logical_basis_bin_codes}.

\section{Problem Setup and Noise Model\label{sec:noise_model}}

In their previous study, the authors of Ref.~\cite{grimsmo_quantum_2020} have assumed ideal preparation of the ancillary state $\ket{+_M}$, ideal \qgate{CROT} gates, as well as an ideal phase measurement: the so called \emph{canonical phase measurement}.
Their simplified noise model suggested that \qcode{cat} and \qcode{bin} codes can go beyond break even by several orders of magnitude.
Nevertheless, implementation of the telecorrection circuit in~\figref{fig:telecorrection} poses many challenges.
A particular challenge lies in performing the logical $\mathcal{M}_{\qgate{X}}$-measurement.
This measurement translates to a phase-estimation problem based on a single copy of the encoded qubit state.
Here we go a step further 
and study the telecorrection circuit in \figref{fig:noise_model_overview}.
This
considers realistic measurement models on the top (data) rail.
These models differ from the canonical phase measurement and are discussed in \secref{ssec:phase_measurements}.
We start in~\secref{ssec:rsb_noise_model} by describing the noise channel. 
    
\subsection{Noise model\label{ssec:rsb_noise_model}}

As in Ref.~\cite{grimsmo_quantum_2020} we consider noisy states that are the result of a noise channel $\mathcal{N}(\hat{\rho})$ which is obtained by integrating the Gorini-Kossakowski-Sudarshan-Lindblad master equation
\begin{align}
   \label{eq:master_equation_loss_dephasing} 
   \pdv{t} \hat{\rho} = \kappa \mathcal{D}[\hat{a}] \hat{\rho} + \kappa_{\phi} \mathcal{D}[\hat{n}] \hat{\rho},
\end{align}
up to some time $\tau$.
Here $\kappa$ and $\kappa_{\phi}$ describe the single photon loss rate and the number dephasing rate, respectively, and $\mathcal{D}(\hat{O}) \hat{\rho} = \hat{O} \hat{\rho} \hat{O}^{\dagger} - \frac12 \hat{O}^{\dagger} \hat{O} \hat{\rho} - \frac12 \hat{\rho} \hat{O}^{\dagger} \hat{O}$ is the Lindblad dissipator.
The resulting quantum channel can be decomposed into a Kraus form, see Ref.~\cite{grimsmo_quantum_2020}.
In all simulations (except for \secref{ssec:comp_rsb_eff}) we consider equal rates $\kappa_{\phi} = \kappa$ even though the noise channels for microwave cavities are typically dominated by photon loss, that is, $\kappa / \kappa_{\phi} \sim 10^2$~\cite{campagne-ibarcq_quantum_2020}.
However, the dispersive coupling to ancillary qubits used for control leads to an increased dephasing rate of the cavity. This can be due to thermal excitations in the qubit resulting in frequency fluctuations of the cavity, for example.
Furthermore, the choice of this relatively large dephasing rate can also be motivated as representing a simplified model for additional noise in the telecorrection circuit
which we describe in the following.

The components that are required to implement the telecorrection circuit in \figref{fig:telecorrection} are:
\begin{enumerate}[label=(\roman*)]
    \item Preparation $\mathcal{P}_{\ket{+_M}}$ of dual-basis code state $\ket{+_M}$; \label{itref:anci_prep}
    \item \qgate{CROT} gates; \label{itref:crot_errs}
    \item Phase measurements for the $\mathcal{M}_{\qgate{X}}$ measurement. \label{itref:faulty_meas}
\end{enumerate}
In principle, any of these three components can be faulty.
However, our numerical analysis focuses on realistically modeling the phase measurements on the data mode (component~\ref{itref:faulty_meas}) because they represent a crucial component of the error correction protocol and are found to quickly dominate all other noise sources for realistic parameters as we show in \secref{sec:num_res_rsb}.

To be explicit, we compare three schemes for the phase measurement: (a) canonical phase measurements (\qmeas{can}), (b) heterodyne detection (\qmeas{het}), and (c) adaptive homodyne detection (\qmeas{ahd}).
An experimental implementation of these schemes will suffer from finite measurement efficiencies $0 \leq \eta < 1$.
In the microwave domain, measurements typically achieve efficiencies in the range of $0.4 \leq \eta \leq 0.75$~\cite{walter_rapid_2017, pfaff_controlled_2017, martin_implementation_2020}.
A practical measurement of an encoded bosonic qubit state must necessarily include mapping the state from a high-$Q$ to an overcoupled low-$Q$ readout mode (with $Q\sim 1/\kappa$ the quality factor) prior to measurement. It is possible that an amplification step can be included prior to releasing the state to a standard microwave chain, thus effectively increasing $\eta$. We do not include any detailed modeling of such a practical scheme here, and simply consider different $\eta$ in the range $0.5 \le \eta \le 1.0$.

We model inefficient measurements by passing the noisy data state $\mathcal{N}(\hat{\rho}_N)$ through a fictional beam splitter before an ideal measurement is performed~\cite{leonhardt_measuring_1997}.
For a measurement with efficiency $\eta$ the noisy state $\mathcal{N}(\hat{\rho}_N)$ transforms as
\begin{align}
    \label{eq:ineff-transform}
    \mathcal{N}(\hat{\rho}_N) \mapsto \Tr_r \left[ \hat{U}_{BS} \left( \mathcal{N}(\hat{\rho}_N) \otimes \ketbra{0}{0}_r \right) \hat{U}_{BS}^{\dagger} \right],
\end{align}
where $r$ is a fictitious reservoir mode (with annihilation operator $\hat{r}$) into which some information is transferred. The beam-splitter transformation is given by
\begin{align}
    \label{eq:beam-splitter-unitary}
    \hat{U}_{BS} = \mathrm{e}^{\arccos(\sqrt{\eta})\left(\hat{a}^{\dagger} \hat{r} - \hat{a} \hat{r}^{\dagger}\right)}.
\end{align}
By performing the trace over the reservoir mode ($\Tr_r[\boldsymbol{\cdot}]$) the information is effectively lost and the output state can be used to model an inefficient measurement.

There are also various types of errors that could occur during the execution of \qgate{CROT} gates implemented through the unitary $\hat{U}_{\qgate{CROT}} = \exp( i \frac{t}{t_g} \frac{\pi}{N M} \hat{n} \otimes \hat{n})$ in time $t_g$.
For example, number-shift errors in one of the modes occurring during the execution of the gate will impose a rotation error (see Eq.~\eqref{eq:crot_error_prop} in the appendix) on the other mode.
The precise rotation angle will be proportional to the time that is left after the number-shift error occurred until the controlled rotation is completed, that is, the specific timing of the number-shift error determines the additional rotation error on the mode. 
However, in any case this rotation error is upper-bounded by $\pi k / N M$, cf. Eq.~\eqref{eq:crot_error_prop}. 
Thus, we can conservatively model number-shift errors during the execution of \qgate{CROT} gates by a higher initial photon loss rate $\kappa \tau$~\footnote{In principle, it is possible to engineer an error transparent Hamiltonian that implements the \qgate{CROT} gate using ancillary two-level systems, see Ref.~\cite{wang_photon-number-dependent_2021}. While the implemented \qgate{CROT} becomes tolerant to photon loss in the bosonic modes, its fidelity is limited by the relaxation of the ancilla systems. Analyzing the performance with error-transparent \qgate{CROT} gates is left for future work.}.
A similar narrative follows for any coherent error that occurs during the execution of the \qgate{CROT} and joint pure loss evolution.
Undesired self-Kerr evolution $(\sim K \hat{n}^2)$ on both modes will be a typical coherent error in any experiment where the controlled rotation is implemented through a cross-Kerr interaction.
The effect of coherent Kerr errors on the code performance were analyzed in Ref.~\cite{albert_performance_2018}. 
Surprisingly, the authors found that small amounts of Kerr $K \lesssim 1$ can benefit \qcode{cat} codes in the case of optimal recovery protocols.
Here we assume that the unitless Kerr parameter $K$ is small, so that the influence of the Kerr-type errors can be approximated by increased dephasing errors~\footnote{See Refs.~\cite{elliott_designing_2018, hillmann_designing_2021} for proposals to effectively cancel self-Kerr couplings.}.
%
Lastly, we do not model errors during state preparation, and we consider perfect encoding $\mathcal{S}_{\qcode{code}} = \hat{S} \boldsymbol{\cdot} \hat{S}^{\dagger}$ 
of the logical information where $\hat{S} = \ketbra{0}_{\qcode{code}} + \ketbra{1}_{\qcode{code}}$ is the encoding operation.

\begin{figure}[h]
    \centering
    \includegraphics{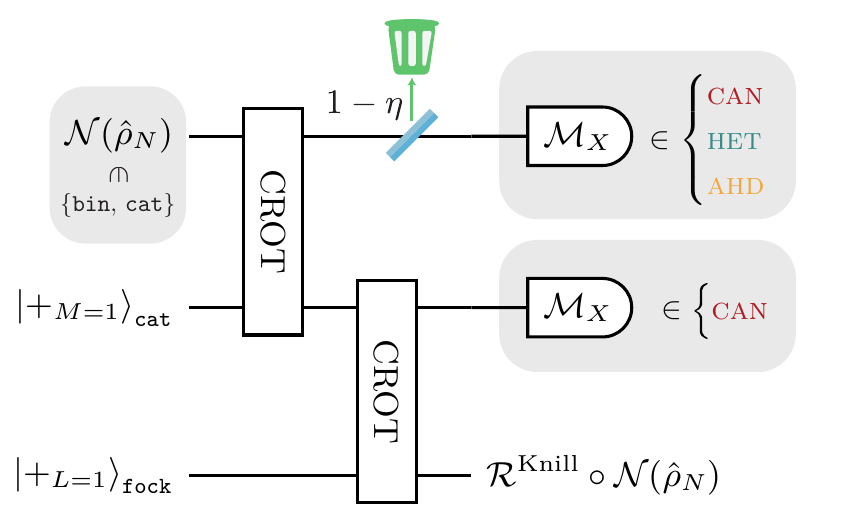}
    \caption{Noise model considered for the telecorrection circuit in \figref{fig:telecorrection}. We restrict our numerical analysis to the two classes of RSB codes discussed in \secref{ssec:explicit_rsb_codes}. The encoded data qubit undergoes a simultaneous loss and dephasing channel $\mathcal{N}(\boldsymbol{\cdot})$ prior to error correction. The noise channel is obtained by integrating the master equation~\eqref{eq:master_equation_loss_dephasing}.
    The measurement on the data (top) rail has efficiency $\eta$, while we model the second measurement on the middle ancilla rail to be near noiseless by choosing an $M=1$ cat code with very large $\alpha$.
    In \secref{ssec:comp_rsb_meas} we analyze the performance of different realizations of the phase measurement for the logical $\mathcal{M}_{\qgate{X}}$ measurement. We simulate the canonical phase measurement (\qmeas{can}), the heterodyne measurement (\qmeas{het}), and the adaptive homodyne scheme (\qmeas{ahd}) for unit efficiency in \secref{ssec:comp_rsb_meas}.
    The results for finite efficiencies in the measurement chain are discussed in \secref{ssec:comp_rsb_eff}.}
    \label{fig:noise_model_overview}
\end{figure}

In particular, with this noise model the complete channel is a logical one, that is, any residual errors on the output state are logical errors \timo{that occur as a result of wrongly decoding at least one of the phase measurements}. 
Therefore, we can set $L=1$ for the (output) bottom rail and use the trivial Fock encoding.
Additionally, we choose a simple $M=1$ \qcode{cat} code with $\alpha = 10$ for the middle ancilla qubit~\cite{grimsmo_quantum_2020} which ensures that measurement errors on the ancilla are negligible compared to all other noise, so that we can consider the ancilla to be essentially noiseless.
\timo{Note that in our noise model where only the data qubit evolves under the noise channel $\mathcal{N}(\hat{\rho})$ there is no incentive to consider different encodings or different realizations of the phase measurement for the first ancilla qubit.
The reason for this is that the ancilla is noise-free and thus one can in principle increase the photon number of the ancilla arbitrarily to obtain any desired accuracy of the phase measurement, that is, any imperfections in the measurement chain can be counteracted by increasing $\overline{n}_{\qcode{code}}$ of the ancilla.
The study of noisy ancilla systems and ancilla measurements is in particular meaningful within a fault-tolerant setting.
This is however beyond the scope of this work and will be subject to a future publication.

}
We illustrate our noise model and initial states in \figref{fig:noise_model_overview}.

\subsection{Different types of phase measurements\label{ssec:phase_measurements}}
\begin{figure*}
    \centering
    \includegraphics{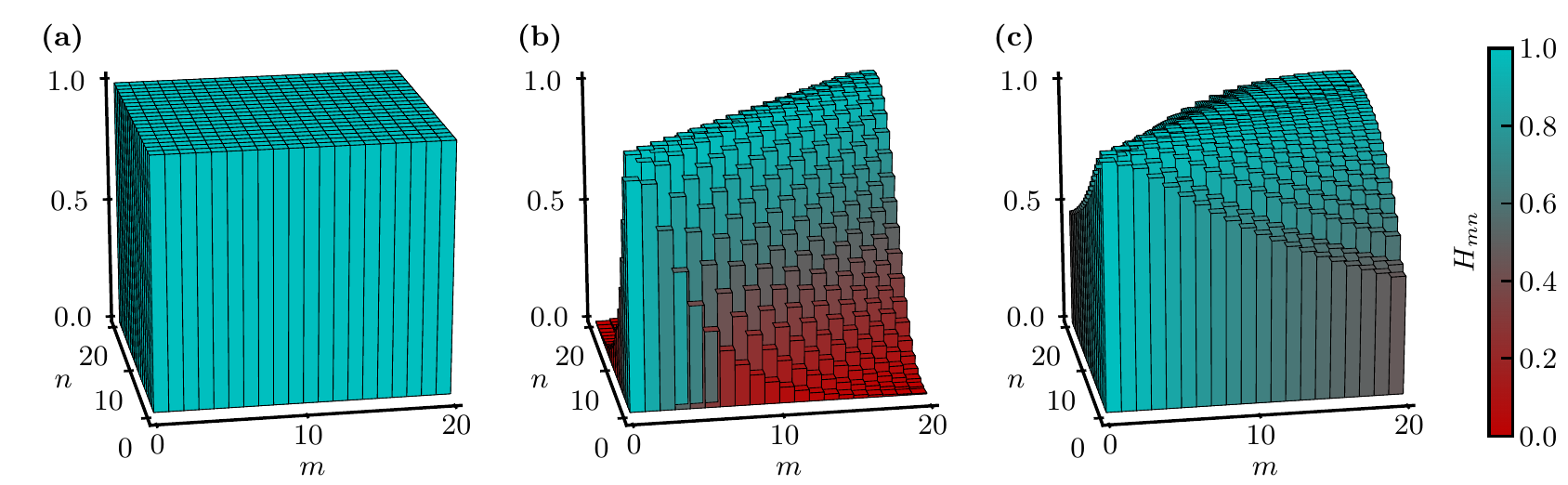}
    \caption{Visualization of the matrix $H$ for different realizations of phase measurements. (a) Canonical phase measurement (\qmeas{can}). (b) Heterodyne detection (\qmeas{het}). (c) Adaptive homodyne detection (\qmeas{ahd}). A good phase measurement is characterized by $H_{mn} = 1\, \forall m, n$, see (a). The comparison of (b) and (c) to  $H^{\qmeas{can}}$ in (a) therefore indicates a clear performance advantage of adaptive homodyne detection over heterodyne measurements for the realization of the logical $\mathcal{M}_{\qgate{X}}$ measurement, \timorevised{see main text for more details.}}
    \label{fig:h_matrices}
\end{figure*}
Defining a suitable observable for the phase in the infinite-dimensional Hilbert space of the harmonic oscillator has been an elusive task~\cite{dirac_quantum_1927, susskind_quantum_1964, pegg_phase_1989}.
Nonetheless, a positive operator valued measurement (POVM) $\hat{F}(\phi)$ for the continuous quantum optical phase $\phi$ can be defined~\cite{leonhardt_canonical_1995}.
Requiring that the POVM elements $\hat{F}(\phi)$ are invariant under phase shifts (phase space rotations) and unbiased, they assume the general form~\cite{leonhardt_canonical_1995, wiseman_adaptive_1998}
\begin{align}
    \label{eq:general_phase_povm}
    \hat{F}(\phi) = \frac{1}{2 \pi} \sum_{n, m = 0}^{\infty} e^{i \phi (m - n)} H_{mn} \ketbra{m}{n},
\end{align}
where $H$ is a positive-semidefinite Hermitian matrix with real and positive entries, and $\phi \in [0,2\pi)$.
Furthermore, the completeness relation $\int_0^{2 \pi} \hat{F}(\phi) \, \mathrm{d}\phi = \mathbb{1}$ implies that $H_{mm} = 1$ for all $m \geq 0$.
The different types of phase measurements are characterized through the off-diagonal elements of $H$ which must be less or equal to one.
For the canonical phase measurement (\qmeas{can}), all elements of $H$ are equal to one.
\timo{
This property allows $\hat{F}^{\qmeas{can}}(\phi)$ to be expressed in terms of pure phase states, that is,
$\hat{F}^{\qmeas{can}}(\phi) = \frac{1}{2 \pi} \ketbra{\phi}$ with
\begin{align}
    \ket{\phi} = \sum_{n=0}^{\infty} e^{i n \phi} \ket{n}.
\end{align}
It can be shown that this property, that is related to the factorization of $H_{mn}$ into $H_{mn} = H_{m} H_{n}^{*}$, is unique to the canonical phase measurement and that any other phase measurement that is described by Eq.~\eqref{eq:general_phase_povm} corresponds to a statistical mixture of canonical phase measurements~\cite{leonhardt_canonical_1995}.
}
However, canonical phase measurements are unphysical~\cite{wiseman_adaptive_1998} and in any realistic phase measurement the off-diagonal elements of $H$ will be less than unity\timo{, reflecting the aforementioned property that they can be understood as statistical mixtures of canonical phase measurements.}

Therefore, a good phase measurement is characterized by a matrix $H$ with elements near the diagonal that are close to unity. 
Intuitively, the deviation from unity of these matrix elements is connected to how much information about the average photon number in addition to the information about the phase $\phi$ is acquired during the measurement.
For the canonical phase measurement no information about the photon distribution is obtained.

Heterodyne measurements are the most convenient way to obtain information about the phase distribution of the quantum state of a microwave field mode. 
This measurement can be realized by performing two homodyne measurements of orthogonal quadratures $\hat{x}_{\theta} = (e^{i \theta} \hat{a} + e^{-i \theta} \hat{a}^{\dagger}) / \sqrt{2}$ and $\hat{x}_{\theta + \pi / 2} = \hat{p}_{\theta}$ simultaneously (at the cost adding noise to both quadratures), see Refs.~\cite{blais_circuit_2021, krantz_quantum_2019} for a comprehensive review.
This realizes a measurement of the coherent state projector $\ketbra{\alpha}$ and the phase information is simply obtained from $\arg(\alpha) = \phi$.
The matrix $H^{\qmeas{het}}$ for heterodyne detection is immediately obtained by expressing $\ketbra{\alpha}$ in the Fock basis, marginalizing the radial component and comparing the result with Eq.~\eqref{eq:general_phase_povm} such that one finds~\cite{wiseman_adaptive_1998}
\begin{align}
    H_{mn}^{\qmeas{het}} = \frac{\Gamma\left( \frac{n + m}{2} + 1 \right)}{\sqrt{\Gamma (n + 1) \Gamma(m + 1)}},
\end{align}
where $\Gamma(x + 1) = x!$ is the Gamma function.
A (theoretically) better alternative to heterodyne detection is adaptive homodyne detection (\qmeas{ahd}).
In contrast to ordinary homodyne detection where the measurement of a single quadrature $\hat{x}_{\theta}$ is performed, in the adaptive scheme the phase $\theta$, which is set by a local oscillator (LO), is continuously updated based on the measurement history, such that the information gain about the photon distribution is minimized.
The theoretical development of adaptive homodyne measurements is contained in Refs.~\cite{wiseman_adaptive_1995, wiseman_adaptive_1997, wiseman_adaptive_1998} and recent experimental results can be found in Ref.~\cite{martin_implementation_2020}.
The analytical expression for the matrix elements of $H^{\qmeas{ahd}}$ is non-trivial and is presented in \appref{app:h_mat_ahd} as given in Ref.~\cite{wiseman_adaptive_1998} for completeness.
Note that the form for $H^{\qmeas{ahd}}$ given there assumes instantaneous feedback of the LO phase $\theta$, meaning that our results will give only an upper bound on the performance of the telecorrection circuit with adaptive homodyne measurements.
The $H$ matrices for the canonical, heterodyne, and adaptive homodyne measurement are visualized in \figref{fig:h_matrices} for $m \leq 20$.
From this figure we expect that adaptive homodyne detection should outperform heterodyne detection.

\begin{figure}[b]
    \centering
    \includegraphics{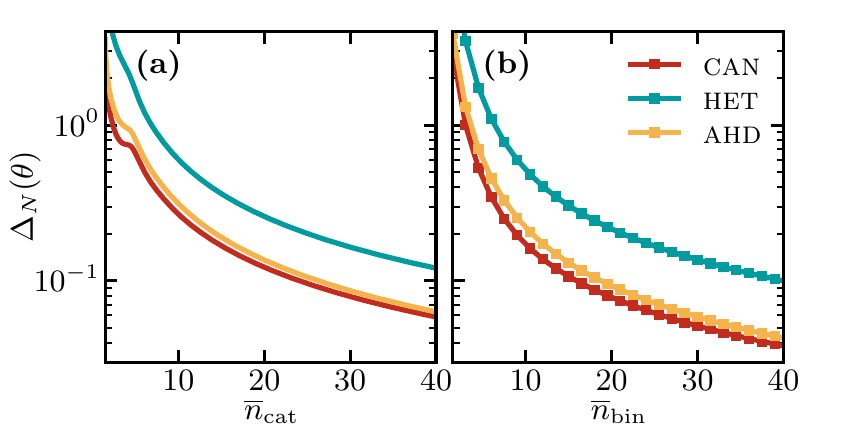}
    \caption{Comparison of the modular phase uncertainty $\Delta_N^{\qmeas{meas}} (\theta)$ for order-$N=3$ number-phase codes and different realizations of the phase measurements. In (a) for \qcode{cat} codes and in (b) for \qcode{bin} codes. \timorevised{
    The continuous lines in (a) emphasize that \qcode{cat} codes are defined for continuous parameter $\alpha$ (with $\alpha^2 = \overline{n}_{\qcode{cat}})$ while markers in (b) emphasize that \qcode{bin} codes are defined only for discrete values of $\overline{n}_{\qcode{bin}}$.}
    Connecting lines in (b) are a guide for the eye.}
    \label{fig:mod_phase_uncertainty}
\end{figure}

To further analyze the differences between the three realizations of phase measurements, we compute the modular phase uncertainty~\cite{grimsmo_quantum_2020}
\begin{align}
    \label{eq:modular_phase_uncertainty}
    \Delta_N^{\mathrm{meas}} (\theta) = \frac{1}{\abs{\left\langle e^{i N \theta }\right\rangle_{\mathrm{meas}}}^2} - 1,
\end{align}
where the mean modular phase $\left\langle e^{i N \theta }\right\rangle_{\mathrm{meas}}$ is related to a measurement scheme by
\newcommand{\phasemeas}{\Omega}
\begin{align}
    \label{eq:mean_mod_phase}
    \left\langle e^{i N \theta }\right\rangle_{\mathrm{meas}} &= \int_0^{2 \pi} e^{i N \theta} \phasemeas(\theta) \dd{\theta} \\
    &= \frac{1}{2} \sum_{k=0}^{\infty} \lvert f_{k N} f_{(k+1)N} H_{k N, (k+1)N}^{\mathrm{meas}} \rvert, \label{eq:mean_mod_phase_1}
\end{align}
with $\phasemeas(\theta) = \Tr[\hat{F}^{\mathrm{meas}}(\theta) \ketbra{+_N}]$, where $\hat{F}^{\mathrm{meas}}$ is the POVM associated to the chosen measurement scheme.
\timorevised{Note that Eq.\eqref{eq:mean_mod_phase_1} shows that in order to minimize the modular phase uncertainty Eq.~\eqref{eq:modular_phase_uncertainty} for an order-$N$ RSB code, the matrix elements $H_{kN, (k+1)N}$ should be close to unity, i.e., $1 -  H_{kN, (k+1)N}  \ll 1$.}
In \figref{fig:mod_phase_uncertainty} 
we study the performance of the different measurements for order-$N=3$ \qcode{cat} and \qcode{bin} codes.
We observe that the canonical phase measurement and adaptive homodyne detection perform similarly, that is, the modular phase uncertainty is dominated by the code states and not by the realization of the measurement.
In contrast, the modular phase uncertainty for order-$N=3$ \qcode{cat} codes is increased by roughly a factor of two for heterodyne measurements, see \sfigref{fig:mod_phase_uncertainty}{a}.
This factor of two is anticipated from the results of Ref.~\cite{wiseman_adaptive_1997} in which the phase uncertainty of coherent states with respect to our measurement schemes is obtained from semi-classical methods.
The gap between adaptive homodyne detection and heterodyne detection widens for \qcode{bin} codes where the modular phase uncertainty of the latter is increased by roughly a factor of $2.5$, see \sfigref{fig:mod_phase_uncertainty}{b}.
%

\section{Numerical Results for RSB codes\label{sec:num_res_rsb}}
In this section we present and discuss our numerical results for the telecorrection circuit subject to the noise model described in \secref{sec:noise_model}.
We begin by comparing the performance of the different phase measurement schemes that were presented in \secref{ssec:phase_measurements} for the ideal case of unit efficiency $\eta = 1$ in the measurement chain.

To asses the performance of the error correction protocol we numerically compute the average gate fidelity~\cite{nielsen_simple_2002, horodecki_general_1999} $\avgFid$ of the resulting quantum channel after applying the most likely logical Pauli correction based on the maximum likelihood decoder [Eq.~\eqref{eq:rsb_decoder}].
\begin{figure*}
    \includegraphics{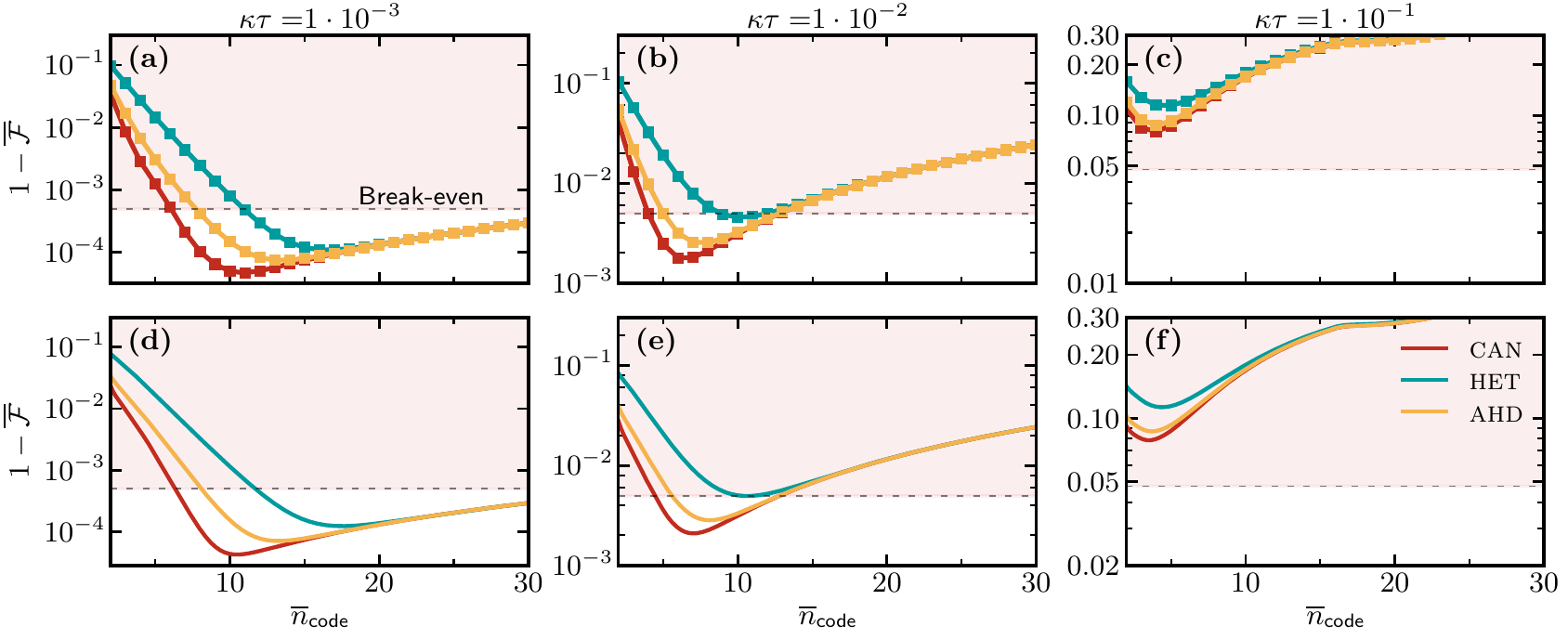}
    \caption{Average gate infidelity for order-$N=2$ \qcode{bin} (top) and \qcode{cat} (bottom) codes as a function of the average code photon number $\overline{n}_{\qcode{code}}$ for different phase measurement schemes.
    The results are obtained by simulating the telecorrection circuit in \figref{fig:noise_model_overview} with equal loss and dephasing strength $\kappa \tau = \kappa_{\phi} \tau$.
    Columns show results for different amounts of noise $\kappa \tau$ before error-correction is performed.
    Error correction is beneficial whenever the infidelity falls below the dashed line that indicates the break-even point, that is, the performance of the trivial Fock encoding without error correction. Note the different scaling of the y-axis through the columns with different $\kappa \tau$.}
    \label{fig:overview_fid_2}
\end{figure*}
\subsection{Comparison of measurement schemes\label{ssec:comp_rsb_meas}}

In all cases we classify whether encoding the logical information and performing error correction by teleportation is advantageous by computing the average gate fidelity for a trivial Fock qubit $\ket{\mu}_{\qcode{triv}} = \ket{\mu} (\mu = 0, 1)$ subject to the same noise channel $\mathcal{N}$ without error correction.
This yields a \emph{break-even point} above which the bosonic encoding is not beneficial. We represent the break-even point by a gray dashed line and shade the region where error correction is not beneficial in pink in all subsequent figures.

We begin by comparing the error-correction potential of the telecorrection circuit for different perfectly-efficient ($\eta = 1$) realizations of phase measurements.
Exemplary results for order-$N=2$ \qcode{bin} and \qcode{cat} codes and selected noise strengths $\kappa \tau$ are presented in \figref{fig:overview_fid_2}.
As observed in Ref.~\cite{grimsmo_quantum_2020}, the results for \qcode{cat} and \qcode{bin} codes are qualitatively similar with a small performance advantage for \qcode{bin} codes.
Furthermore, we find that there is a hierarchy in performance between the different realizations of phase measurements.
The heterodyne measurement achieves the lowest average gate fidelity, followed by the adaptive homodyne measurement and the canonical phase measurement.
This hierarchy was anticipated from the discussion in \secref{ssec:phase_measurements} and \figref{fig:h_matrices} therein.
Additionally, for any measurement scheme and noise strength there exists a code characterized by its average photon number $\overline{n}_{\qcode{code}}$, i.e., a sweet spot, that maximizes the average gate fidelity.
We can formulate the main takeaway message from \figref{fig:overview_fid_2} as follows:
by replacing the canonical phase measurement of the encoded data qubit with a physically realizable measurement \qmeas{het} the break-even potential for RSB codes using error-correction by teleportation (\figref{fig:telecorrection}) reduces significantly, while with \qmeas{ahd} the reduction is relatively small.
Nonetheless, using realizable measurement models, RSB codes can still perform better than break-even, even though this comes at the cost of code words with increased average photon number $\overline{n}_{\qcode{code}}$ in comparison to the results obtained with \qmeas{can} measurements.
Increasing the code size comes at the cost of increasing the susceptibility towards nonlinear coherent error sources, such as self- and cross-Kerr interactions.
The effects of those are not included into this numerical analysis and are beyond the scope of this work.

\begin{figure}[t]
    \includegraphics{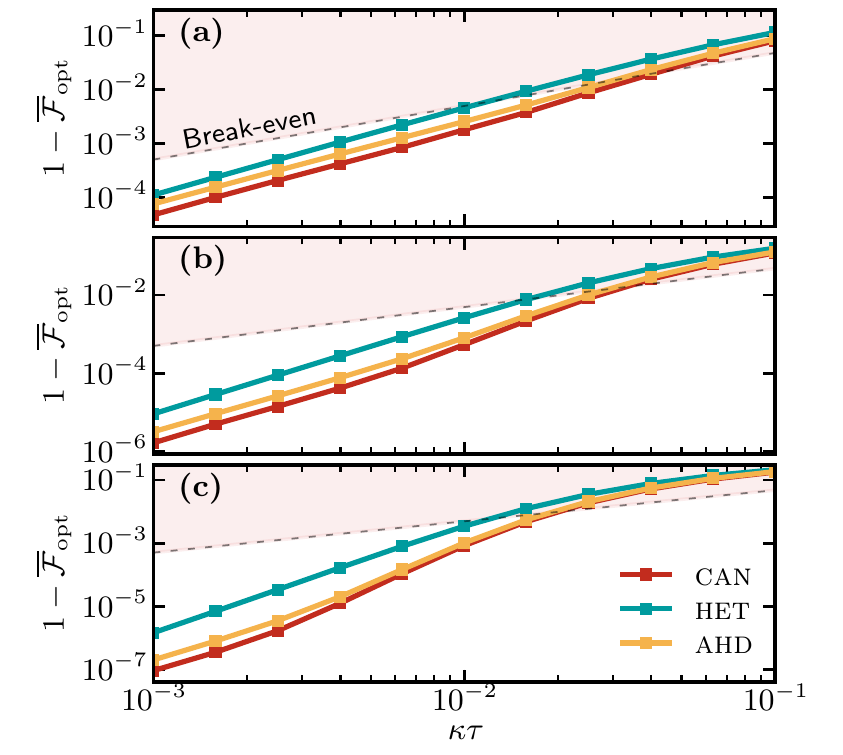}
    \caption{Maximum average gate infidelity as a function of noise strength $\kappa \tau$ for \qcode{bin} codes, and different realizations of the phase measurement.
    For each value of $\kappa \tau$, the code with the optimal average photon number $\overline{n}_{\qcode{code}}$ that maximizes $1-\avgFid$ is chosen.
    Panels (a), (b) and (c) show the results for the order-$N=2$, $N=3$ and $N=4$ \qcode{bin} codes, respectively.}
    \label{fig:break_even_bin}
\end{figure}

\begin{figure*}
    \centering
    \includegraphics{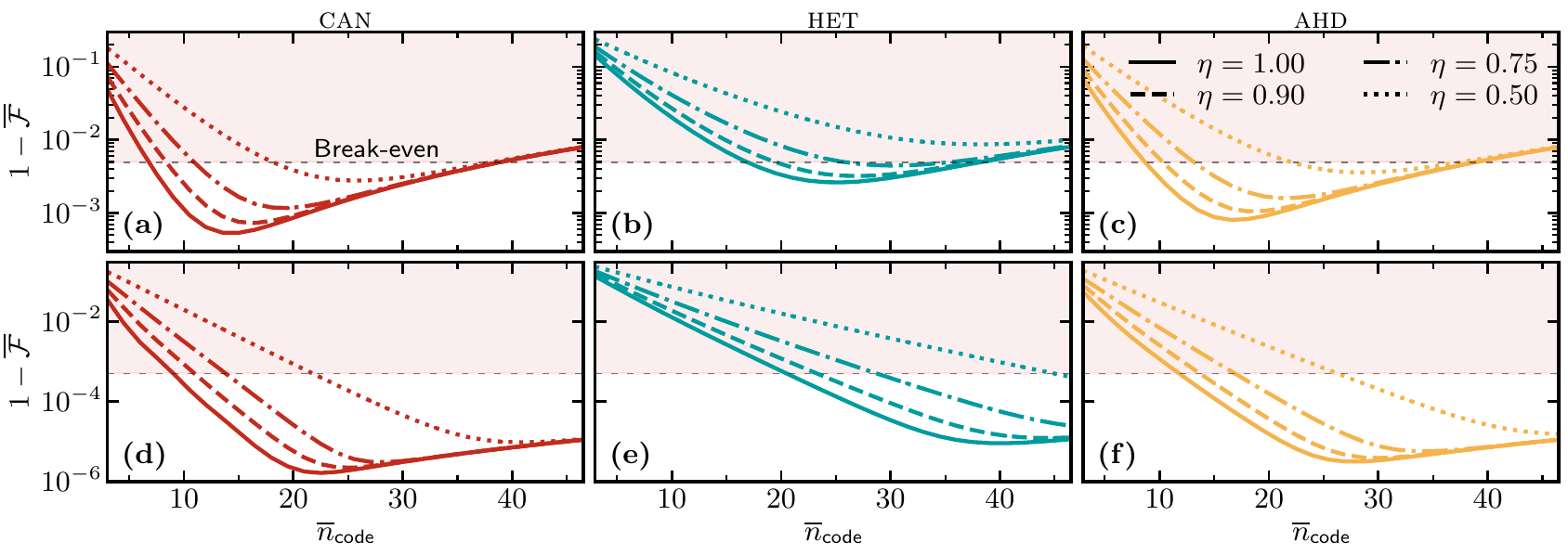}
    \caption{Average gate infidelity for the telecorrection circuit (\figref{fig:telecorrection}) with inefficient measurement models. Before error-correction is applied the initial state is subject to noise of total strength $\kappa \tau = \kappa_{\phi} \tau = 1 \cdot 10^{-2}$ in panels (a)-(c) (top row) and $\kappa \tau = \kappa_{\phi} \tau = 1 \cdot 10^{-3}$ in panels (d)-(f) (bottom row).
    Each column shows a different realization of the phase measurement.
    The figure shows the results for the order-$N=3$ \qcode{bin} codes which are defined only for discrete  $\overline{n}_{\qcode{code}}$ values. Continuous lines are shown to improve readability.
    Error correction with RSB codes is beneficial whenever the infidelity falls below the dashed line that indicates the break-even point.}
    \label{fig:ineff_fid_bin_N=3}
\end{figure*}
To further analyze the break-even potential for realistic measurement models we compute the average gate fidelity for different rotation-symmetry $N$.
\figref{fig:break_even_bin} shows the average gate infidelity $1-\avgFid_{\mathrm{opt}}$ for the \qcode{bin} code with the optimal $\overline{n}_{\qcode{code}}$ as a function of the total noise strength $\kappa \tau = \kappa_{\phi} \tau$.
We can identify two central points from the shown results which are qualitatively similar for \qcode{bin} and \qcode{cat} codes: 
first, observe that $1-\avgFid_{\mathrm{opt}}$ drops proportionally to the total noise $\kappa \tau$ for sufficiently small values $\kappa \tau$.
Remarkably, the proportionality constant is approximately the same independently of the concrete realization of the phase measurement scheme and the functional dependence only differs by a measurement dependent offset.
Secondly, the gap between the heterodyne and adaptive homodyne measurement schemes increases with the order $N$ of the code. 
\timo{It is unclear whether} this trend continues arbitrarily. 
However, our analysis is limited to order-$N \leq 4$ codes due to the large Fock-space needed to reach the required numerical accuracy for $N \geq 5$, which makes simulation difficult.
It should be emphasized, however, that \figref{fig:break_even_bin} obscures the fact that the similar scaling and performance of all measurement schemes comes at the cost of larger code states as discussed above.

\subsection{Finite detection efficiencies\label{ssec:comp_rsb_eff}}
Up to this point we have limited our numerical analysis to perfectly efficient ($\eta = 1$) measurements.
However, the case of finite measurement efficiencies $0 \leq \eta < 1$ is relevant in the microwave domain where measurement efficiencies 
are typically limited to the range $0.5 \leq \eta \leq 0.75$~\cite{walter_rapid_2017,martin_implementation_2020, pfaff_controlled_2017}.
As mentioned above, one might hope that $\eta$ can be effectively increased by a cleverly designed measurement schemes that amplifies the signal \emph{prior} to releasing it to the measurement chain.
We analyze the impact of finite measurement efficiencies on the average gate fidelity for different realizations of the phase measurement for order-$N=3$ \qcode{bin} codes in the panels of \figref{fig:ineff_fid_bin_N=3}.
In this figure we exemplify our findings for two different noise strengths $\kappa \tau =  \kappa_{\phi} \tau =1 \cdot 10^{-2}$ (top row) and $\kappa \tau =  \kappa_{\phi} \tau =1 \cdot 10^{-3}$ (bottom row).
It should be emphasized that the observations described below are qualitatively similar for all codes and noise strengths that we have studied.
We point out a few observations from \figref{fig:ineff_fid_bin_N=3} that are worth mentioning.
First, the response to a finite efficiency in the measurement realization for the \qmeas{can} and \qmeas{ahd} measurements is similar.
More importantly, the telecorrection circuit for RSB codes is not extremely sensitive to finite efficiencies, that is, for high measurement efficiencies ($\eta = 0.9$) the average gate fidelity does not reduce abruptly in comparison to the ideal case ($\eta=1$).
However, by reducing the efficiencies further to values of $\eta = 0.75$ and $\eta = 0.5$, significant gaps to the ideal case arise, reducing the break-even potential for approximate number-phase codes.
Nevertheless, it is in principle still possible to reach break-even if the noise rate is low enough, see in particular the results for \qmeas{het} measurements with $\eta = 0.5$ in \sfigref{fig:ineff_fid_bin_N=3}{e}.
The results also highlight a significant potential for improvement that comes with the development of advanced phase measurement techniques such as adaptive homodyne detection over standard heterodyne detection [cf. \sfigref{fig:ineff_fid_bin_N=3}{b} and \sfigref{fig:ineff_fid_bin_N=3}{c}] in order to increase the break-even potential of RSB codes.

The results can be understood qualitatively by realizing the effect of inefficiencies on the primitive state of \qcode{cat} codes which is the coherent state $\ket{\alpha}$.
In that case, the effect of Eq.~\eqref{eq:ineff-transform} is to reduce the amplitude $\alpha \mapsto \sqrt{\eta} \alpha$ of the primitive state without increasing its uncertainty $\bra{\alpha}(\hat{x} - \langle \hat{x} \rangle)^2\ket{\alpha} = 1/2$. As a result, measurements with high efficiencies do not alter $1 - \overline{\mathcal{F}}$ significantly~\footnote{We have checked that for the codes and the error-correction circuit here considered it is not advantageous to concatenate the loss channel Eq.~\eqref{eq:ineff-transform} with an amplification channel as typically done for GKP codes~\cite[see, e.g.,][and also Sec.~\ref{sssec:twirling_approx}]{albert_performance_2018}}.
Only for large losses in the measurement chain does this effect become significant, effectively increasing the modular phase uncertainty $\Delta_{N}(\theta)$.

The maximum average gate fidelity at the code sweet spot for finite-efficiency measurements behaves qualitatively similar to the instance with unit efficiency in \figref{fig:break_even_bin}.
The quantitative difference is a reduction in the break-even potential.
This reduction is small for $\eta = 0.9$ and becomes noticeable for $\eta = 0.75$ and $\eta = 0.5$.
\section{Gottesman-Kitaev-Preskill codes\label{sec:gkp_comparison}}
Another candidate towards fault-tolerance with bosonic codes was developed by Gottesman, Kitaev, and Preskill (GKP)~\cite{gottesman_encoding_2001} to protect against (small) shifts of the phase space variables $x$ and $p$.
It therefore differs 
from the
RSB codes introduced above, which are designed to protect against loss, gain, and dephasing errors.
An earlier study has shown a theoretical performance advantage
of GKP codes over codes that belong 
to the RSB class~\cite{albert_performance_2018} for the case of a pure loss noise channel.
This theoretical advantage motivated further research towards practical computation with GKP codes~\cite{tzitrin_progress_2020, hastrup_unsuitability_2021}.
Here, we contribute to these developments 
by studying the performance of teleportation-based error correction
with GKP codes using realistic measurement models and  comparing these results with those obtained in \secref{sec:num_res_rsb} for RSB codes.
To assess the performance we have carried out exact Fock space simulations as well as calculations using
an approximate analytical model that is based on the twirling approximation.
This model is commonly used for estimating fault-tolerant thresholds of GKP codes concatenated with stabilizer codes~\cite{noh_fault-tolerant_2020, menicucci_fault-tolerant_2014, fukui_all-optical_2021, vuillot_quantum_2019, larsen_fault-tolerant_2021}.
We note that teleportation-based error correction for GKP codes with the Knill $C_4 / C_6$ scheme was analyzed in Ref.~\cite{fukui_analog_2017} using the twirling approximation.

We begin by reviewing relevant notation for GKP codes before describing the analyzed noise model and error correction circuit.


\subsection{Background}
The GKP encoding is a continuous-variable stabilizer code defined via the two commuting stabilizers~\cite{gottesman_encoding_2001}
\begin{subequations}
\label{eq:gkp_stabilizers}
\begin{align}
    \hat{S}_x &= e^{i 2 \sqrt{\pi} \hat{x}} = \qgate{Z}^2, \label{eq:gkp_x_stabilizer} \\
    \hat{S}_p &= e^{-i 2 \sqrt{\pi} \hat{p}} = \qgate{X}^2. \label{eq:gkp_p_stabilizer}
\end{align}
\end{subequations}
The two stabilizers in Eq.~\eqref{eq:gkp_stabilizers} act as translation operators in phase space so that the ideal code states $\ket{0}_{\qcode{gkp}}$ ($\ket{+}_{\qcode{gkp}}$) and $\ket{1}_{\qcode{gkp}}$ ($\ket{-}_{\qcode{gkp}}$) become infinite superpositions of position (momentum) eigenstates located at even and odd multiples of $\sqrt{\pi}$, respectively.
That is, the unnormalized ideal code states take the following form
\begin{align}
    \label{eq:ideal_gkp_code}
    \ket{\mu}_{\qcode{gkp}} = \sum_{n=-\infty}^{\infty} \ket{(2n + \mu) \sqrt{\pi}}_x,
\end{align}
for $\mu = 0, 1$ and $\ket{x}_x$ denotes the position eigenstate satisfying $\hat{x} \ket{x}_x = x \ket{x}_x$.
The dual-basis code states $\ket{\pm}_{\qcode{gkp}}$ are obtained from Eq.~\eqref{eq:ideal_gkp_code} through the Fourier relation $\hat p = \exp{i \frac{\pi}{2} \hat{a}^{\dagger} \hat{a}} \hat{x} \exp{-i \frac{\pi}{2} \hat{a}^{\dagger} \hat{a}}$~\cite{gottesman_encoding_2001}, that is, 
\begin{align}
    \label{eq:ideal_gkp_dual}
    \ket{\pm}_{\qcode{gkp}} = \sum_{n = - \infty}^{\infty} \ket{[2 n + (1 \pm 1)/2] \sqrt{\pi}}_p
\end{align}
where $\ket{p}_p$ denotes a momentum eigenstate.
The ideal code states allow for correcting a continuous set of displacements with shifts of at most $\sqrt{\pi} / 2$ along both quadratures.
These shift errors can be corrected either by Steane-type or Knill-type error correction through measurement of the stabilizers and suitable correction operations~\cite{gottesman_encoding_2001, glancy_error_2006}.
We consider Knill-type error correction in the following in more detail to compare it to the telecorrection circuit used for arbitrary RSB codes discussed in \secref{sec:noise_model}. 

\subsection{Noise model and error correction circuit\label{ssec:gkp_noise_correction}}
The ideal GKP code states $\ket{\mu}_{\qcode{gkp}}$ are infinite energy states and therefore unphysical.
In practice, the ideal, infinitely squeezed code states need to be approximated by states with finite squeezing.
This can be achieved, for example, by regularizing the code states as~\cite{menicucci_fault-tolerant_2014, royer_stabilization_2020}
\begin{align}
    \label{eq:regularized_code_words}
    \ket{\mu_{\delta}}_{\qcode{gkp}} \propto \hat{E}_{\delta} \ket{\mu}_{\qcode{gkp}},
\end{align}
where $\hat{E}_{\delta}$ is an envelope operator that suppresses high-energy contributions.
Here we choose
\begin{align}
    \label{eq:envelope_op}
    \hat{E}_{\delta} = \mathrm{e}^{- \delta^2 \left(\hat{n} + \frac{1}{2}\right)},
\end{align}
with regularization parameter $\delta$ that characterizes the squeezing of each individual peak.
It is noteworthy that other approximations of GKP states are possible~\cite[see also Eq.~\eqref{eq-app:gkp_finite_energy} in \appref{app:numerical_details}]{albert_performance_2018, gottesman_encoding_2001, motes_encoding_2017}, however, these approximations have been shown to be equivalent to the regularization through the envelope operator in Eq.~\eqref{eq:envelope_op}~\cite{matsuura_equivalence_2020}.

\begin{figure}[t]
    \centering
    \begin{quantikz}
    \lstick{$\ket{\psi}_{\qcode{gkp}}$} & \qw & \gate{E_{\delta}} & \gate{\mathcal{N}} & \ctrl{1} &  \gate{\mathcal{A}_G \circ \mathcal{L}_{\eta}} &\meterD{\hat{p}} \rstick{$p_1$} \\
    \lstick{$\ket{+}_{\qcode{gkp}}$} & \ctrl{1} & \gate{E_{\delta}} & \qw & \control{} &  \meterD{\hat{p}} \rstick{$p_2$} \\
    \lstick{$\ket{+}_{\qcode{gkp}}$} & \control{} & \qw & \gate{\mathcal{R}^{\mathrm{Knill}}} & \qw  \rstick{$\hat{\rho}_{\mathrm{out}}$}
    \end{quantikz}
    \caption{Teleportation based error-correction circuit for GKP codes with the type and location of noise terms we are considering.
    We assume that the ancilla states are regularized through $E_{\delta}$ only after they are entangled through the \qgate{CZ} operation such that this circuit implements a logical channel of the form Eq.~\eqref{eq:rsb_knill_ec_out}.
    The regularized input state is then subject to a simultaneous loss and dephasing channel $\mathcal{N}$ that is obtained from the master equation~\eqref{eq:master_equation_loss_dephasing}.
    The measurement model assumes standard homodyne detection and we model finite efficiencies $0 \leq \eta \leq 1$ either by a pure loss channel $\mathcal{L}_{\eta}$ or by random displacement channel $\mathcal{N}_{\sigma}$ that is obtained by concatenating $\mathcal{L}_{\eta}$ with an amplification channel $\mathcal{A}_G$ with gain $G = 1/\eta$, see main text for details. \timo{ The recovery map $\mathcal{R}^{\mathrm{Knill}}$ depends on the measurement outcome $\vec{p} = (p_1, p_2)$, cf. Eq.~\eqref{eq:rsb_knill_ec_out}.}}
    \label{fig:gkp_telecorrection}
\end{figure}
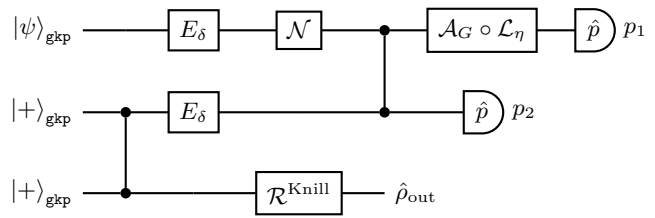

The Knill-type error correction circuit that we are considering for GKP codes is shown in \figref{fig:gkp_telecorrection}.
The error correction capabilities of this circuit are analogous to the one in \figref{fig:telecorrection} for RSB codes.
In order to obtain a logical channel from the circuit, we must slightly alter the noise model that was considered in \secref{sec:noise_model} for RSB codes.
The reason is that the regularization in Eq.~\eqref{eq:envelope_op} leads to a propagation of errors through the entangling $\qgate{CZ} = \exp( i \hat{x}_1 \otimes \hat{x}_2)$ gate to the output state $\hat{\rho}_{\mathrm{out}}$.
To render the resulting channel logical, we therefore assume that the ancilla states are regularized only \emph{after} the \qgate{CZ} gate between them has been performed.
This assumption is made so that we can compare the results of this section to the previous ones.
For the data rail the noise model is unchanged, that is, the simultaneous loss and dephasing channel $\mathcal{N}$ acts on the regularized  encoded qubit state $\hat{E}_{\delta} \ket{\psi}_{\qcode{gkp}}$.
The logical $\mathcal{M}_{\qgate{X}}$ measurement in this case corresponds to a homodyne measurement along the $p$-quadrature.
Hence, the resulting output state $\hat{\rho}_{\mathrm{out}}$ is given by an expression that is analogous to Eq.~\eqref{eq:rsb_knill_ec_out}.
The measurement outcome is decoded by closest-integer decoding, that is, by rounding each of the measurement results $\vec{p} = (p_1, p_2) \mod 2\sqrt{\pi} $ to the nearest integers $0$ or $1$.

In the spirit of the noise model for RSB codes we model finite-efficiency ($0 < \eta < 1$) homodyne measurements by a lossy beam splitter before an ideal measurement, see Eq.~\eqref{eq:ineff-transform}.
Here and in \figref{fig:gkp_telecorrection} we denote the loss channel corresponding to a measurement with efficiency $\eta$ by $\mathcal{L}_{\eta}$.
Because GKP codes are designed to protect against shift errors, it is common~\cite{gottesman_encoding_2001, albert_performance_2018, noh_fault-tolerant_2020, fukui_all-optical_2021} to concatenate the loss channel $\mathcal{L}_{\eta}$ with an amplification channel $\mathcal{A}_G$ with gain $G = 1/\eta$ that exactly compensates for the losses and transforms $\mathcal{L}_{\eta}$ into a random displacement channel $\mathcal{N}_{\sigma}= \mathcal{A}_{1/\eta} \circ \mathcal{L}_{\eta}$ with variance $\sigma^2 = (1-\eta)/\eta$~\cite{ivan_operator-sum_2011, noh_quantum_2019}, where $\mathcal{N}_{\sigma}$ is given by,
\begin{align}
    \label{eq:displacement_channel}
    \mathcal{N}_{\sigma} (\boldsymbol{\cdot}) = \frac{1}{\pi \sigma^2} \int \exp[- \frac{\lvert \alpha \rvert^2}{ \sigma^2}] \hat{D}(\alpha) \boldsymbol{\cdot} \hat{D}^{\dagger}(\alpha) \dd[2]{\alpha}.
\end{align}
We compare both of the above mentioned approaches (i.e., with and without $\mathcal A_G$) in \secref{ssec:gkp_num_res}.

\subsubsection{Analytical model through twirling approximation\label{sssec:twirling_approx}}
Fock space simulations of the error correction circuit in \figref{fig:gkp_telecorrection} are quickly unmanageable for $\delta \ll 1$ even though the average photon number $\overline{n}_{\qcode{gkp}}$ is relatively small compared to the Fock space truncation $N_{\mathrm{trunc}}$ required.
This can be understood from the special structure of the code states $\ket{\mu_{\delta}}_{\qcode{gkp}}$ in phase space which have peaks far from the origin and thus require Fock states $\ket{n}$ with large $n$, even though the peaks have low weight.

To circumvent this issue, an analytical tractable model can be devised by realizing that the envelope operator can be expressed in terms of a coherent superposition of displacements~\cite{matsuura_equivalence_2020, noh_fault-tolerant_2020}.
By applying uniformly random displacements the coherent superposition can be converted to an incoherent mixture of Gaussian displacements~\cite{noh_fault-tolerant_2020}.
This conversion is known as state-twirling, but it is unphysical as it requires the application of displacements $\hat{D}(\alpha)$ with infinite energy ($\alpha \to \infty$)~\cite{conrad_twirling_2021}.
Still, state-twirling is useful for GKP codes to derive an approximate analytic model as it reduces the problem to tracking the variance $\sigma_{\qcode{gkp}}^2$ of a single Gaussian distribution with zero mean if all other noise channels can be converted by twirling to random displacement channels with variance $\sigma_{\mathrm{noise}}^2$ as well.
This is the case for typical noise processes such as single photon loss and gain, but it is unknown if this is also possible for the the number dephasing channel.

By calculating how the quadrature noise of the data and ancilla states is transformed by the circuit \figref{fig:gkp_telecorrection} one finds that the probability for decoding one of the homodyne measurement outcomes wrongly is given by
\begin{align}
    \label{eq:p_err_gkp}
    P_{\mathrm{err}}(\sigma_{\mathrm{data}}, \sigma_{\mathrm{anci}}) = 1 - p_{\mathrm{succ}}(\sigma_{\mathrm{data}}) p_{\mathrm{succ}}( \sigma_{\mathrm{anci}}),
\end{align}
where the individual success probabilities $p_{\mathrm{succ}}(\sigma)$ are given by
\begin{align}
    \label{eq:p_succ_gkp}
    p_{\mathrm{succ}}(\sigma) = \frac{1}{\sqrt{2 \pi \sigma^2}} \sum_{n \in \mathbb{Z}} \int_{(2 n - 1/2) \sqrt{\pi}}^{(2 n + 1/2) \sqrt{\pi}} e^{-\frac{1}{2} \frac{z^2}{\sigma^2}} \dd{z}.
\end{align}
In the presence of other noise channels that are characterized by a Gaussian probability distribution the effective variance can be calculated as $ \sigma_{\mathrm{eff}}^2 = \sum_{i \in I} \sigma_i^2$ with $I$ a set of indices that represents all possible noise sources.
In particular, for the model in \figref{fig:gkp_telecorrection} we obtain,
\begin{subequations}
\label{eq:gkp_noise_variance}
\begin{align}
    \sigma_{\mathrm{eff, data}}^2 = \frac{1}{2} \left( \delta_{\mathrm{data}}^2 + \delta_{\mathrm{anci}}^2 \right) + \frac{1 - \eta}{\eta}, \label{eq:gkp_data_variance} \\
     \sigma_{\mathrm{eff, anci}}^2 = \frac{1}{2} \left( \delta_{\mathrm{data}}^2 + \delta_{\mathrm{anci}}^2 \right), \label{eq:gkp_anci_variance}
\end{align}
\end{subequations}
where the factor $1/2$ appears due to the relation $\delta^2 = 2 \sigma^2_{\qcode{gkp}}$ and the last term in Eq.~\eqref{eq:gkp_data_variance} is the result of modeling the finite efficiency measurements by a random displacement channel before the ideal homodyne measurement.
For more details see, e.g.,~\cite{noh_fault-tolerant_2020, noh_low-overhead_2022, fukui_high-threshold_2019, fukui_all-optical_2021}.

\subsection{Numerical results\label{ssec:gkp_num_res}}
In this section we present our numerical results for the noise model and error correction circuit described in \secref{ssec:gkp_noise_correction} and compare them to the analytical results obtained through the twirling approximation if applicable.
To this end, we numerically compute the entanglement fidelity $\entFid$~\cite{schumacher_quantum_2010} while assuming perfect encoding and decoding of the logical information as defined in \secref{ssec:rsb_noise_model}.
However, note that for a $d$-level system $\entFid$ and $\avgFid$ are related through the relation~\cite{nielsen_simple_2002}
\begin{align}
    \label{eq:rel_avg_ent_fid}
    \avgFid = \frac{d \entFid + 1}{d + 1},
\end{align}
with $d=2$ for the case of the logical qubit-to-qubit channel considered here, so that a quantitative comparison is possible.
Furthermore, it can be shown (see \appref{app:entanglement_fid} for details) that the logical error probability $P_{\mathrm{err}}$ obtained from the twirling approximation can be related to the entanglement fidelity $\entFid$, and therefore also to the average gate fidelity $\avgFid$.
Simulation details are described in \appref{app:numerical_details}.

\subsubsection{Comparison of analytical and numerical results\label{sssec:comp_ana_num}}
\begin{figure}[t]
    \centering
    \includegraphics{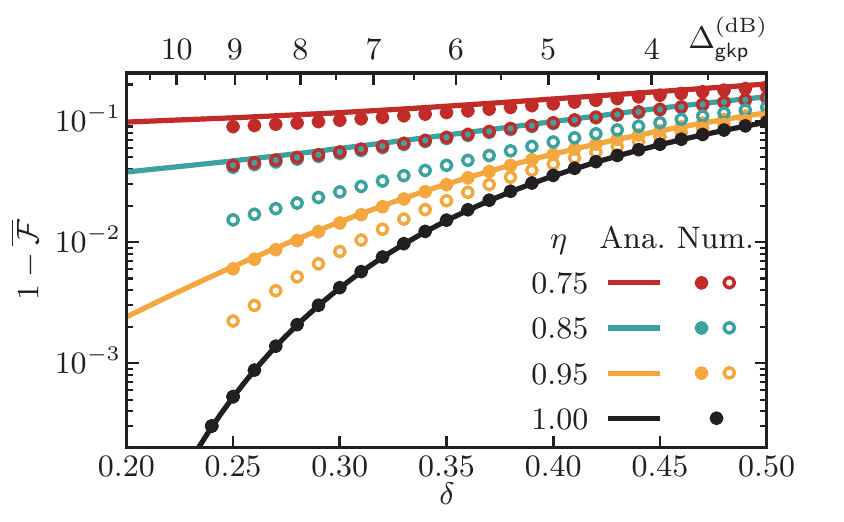}
    \caption{Comparison of the average gate infidelity $1-\avgFid$ 
    of the telecorrection circuit in \figref{fig:gkp_telecorrection} in the absence of the loss and dephasing channel $\mathcal{N}$,
    calculated through numerical Fock space simulations (filled dots) and through the twirling approximation (solid lines) as a function of the regularization parameter $\delta = \delta_{\mathrm{data}} = \delta_{\mathrm{anci}}$ (squeezing $\Delta^{(\mathrm{dB})}_{\qcode{gkp}}$).
    We show results for unit-efficiency homodyne detection as well as finite detection efficiency $0 \leq \eta < 1$ modeled as random displacement channel $\mathcal{N}_{\sigma} = \mathcal{A}_{1/\eta} \circ \mathcal{L}_{\eta}$, see main text for details.
    Additionally, open dots show $1 -\avgFid$ when finite efficiencies are modeled by a pure loss channel $\mathcal{L}_{\eta}$ and the decision boundaries of the closest-integer decoding are rescaled by $\sqrt{\eta}$.
    Numerical results are obtained with Fock space truncation $N_{\mathrm{trunc}}=250$ which is sufficient up to $\delta \approx 0.225$. 
    }
    \label{fig:rescaled_gkp_fock_twirling}
\end{figure}

We begin by considering the case where the only noise sources are finite squeezing $\delta$ and finite measurement efficiency $\eta$ in order to establish a comparison between the exact Fock space simulation of the regularized GKP states and the twirling approximation.
The results for this case are summarized in \figref{fig:rescaled_gkp_fock_twirling}, which shows the average gate infidelity $1-\avgFid$ as a function of $\delta = \delta_{\mathrm{data}} = \delta_{\mathrm{anci}}$ for various measurement efficiencies $\eta$.
We observe an excellent agreement between the Fock space simulation and the twirling approximation for $\eta = 1$ up to $\delta \approx 0.225$ when the Fock space truncation $N_{\mathrm{trunc}}=250$ becomes insufficient.
This corresponds to a squeezing level of approximately $\Delta_{\qcode{gkp}}^{(\mathrm{dB})} = 10 \log_{10}\left( \frac{1}{2 \delta^2} \right) \approx 10 \, \mathrm{dB}$.
For $0.75 \leq \eta \leq 0.95$ the agreement is reduced with $\eta$, but the twirling approximation still presents a relatively tight upper bound on the achievable infidelity.
Notice, however, that for $\eta = 0.75$ and $\delta < 0.25$ the infidelity $1-\avgFid$ has almost reached a plateau since the noise due to the finite efficiency measurement now dominates.

Our results therefore strengthen the recent results for fault-tolerance thresholds obtained for GKP-surface codes in Ref.~\cite{noh_low-overhead_2022} in the sense that the twirling approximation results in tight bounds for the achievable physical error rates if all noise besides the state regularization can be described by a Gaussian random displacement channel $\mathcal{N}_{\sigma}$.
However, our results also indicate that the average gate infidelity $1 -\avgFid$, and therefore also the logical error rate $P_{\mathrm{err}}$, are highly sensitive to finite measurement efficiencies.
For example, a measurement efficiency of $95 \, \%$ increases the achievable infidelity at $\delta = 0.25$ by an order of magnitude in comparison to the ideal case $\eta = 1$.
This agrees well with the results of Ref.~\cite{fukui_high-threshold_2019} which present $\eta \approx 92.2 \,\%$ as a lower bound on the necessary efficiency for topologically protected measurement based quantum computing with GKP codes.
%

In \figref{fig:rescaled_gkp_fock_twirling} we also show $1 - \avgFid$ for the case when finite efficiencies are modeled by a pure loss channel $\mathcal{L}_{\eta}$, that is, losses are not compensated through amplification.
This noise model does not admit a description within the analytical model introduced above and we are restricted to numerical Fock space simulations.
For this case we have rescaled the decision boundaries of the closest-integer decoding by $\sqrt{\eta}$ to amount for the losses.
We found this improves the decoding significantly over the standard decision boundaries.
Similar ideas were discussed in Ref.~\cite{fukui_all-optical_2021} and are denoted as classical computer (cc) amplification.
We can draw the following conclusions from \figref{fig:rescaled_gkp_fock_twirling}.
First, our results show that it is not beneficial to perform amplification through a quantum-limited amplification channel $\mathcal{A}_{1/\eta}$ that exactly compensates for the losses.
The reason is that the photon loss channel $\mathcal{L}_{\eta}$ increases the variance of each peak and moves each peak by a factor $\sqrt{\eta}$ closer to the origin.
Effectively, only the peak variance increases when the decision boundaries are rescaled.
Ref.~\cite{fukui_all-optical_2021} showed that the noise added by this procedure is only half as large as that of the random displacement channel $\mathcal{N}_{\sigma} = \mathcal{A}_{1/\eta} \circ \mathcal{L}_{\eta}$, that is, the effective variance is $\sigma^2_{\mathrm{cc}} = (1 - \eta) / 2 \eta$.
Indeed, we have checked that our exact Fock space simulation of a regularized state that is subject to the loss channel $\mathcal{L}_{\eta}$ before the ideal homodyne measurement shows excellent agreement with the results obtained from the twirling approximation when $\mathcal{L}_{\eta}$ is modeled by an effective displacement channel with effective variance $\sigma^2_{\mathrm{cc}}$.
Furthermore, comparing \figref{fig:ineff_fid_bin_N=3} in \secref{sec:num_res_rsb} and \figref{fig:rescaled_gkp_fock_twirling} here, our results demonstrate that GKP codes are much more sensitive to finite detection efficiencies than RSB codes.

\subsubsection{Comparison to RSB codes \label{sssec:rsb_gkp_comp}}
We now compare the performance of GKP and RSB subject to the same noise channel $\mathcal{N}$ that is obtained by integrating Eq.~\eqref{eq:master_equation_loss_dephasing} and describes simultaneous loss and dephasing.

\begin{figure*}[!t]
    \centering
    \includegraphics{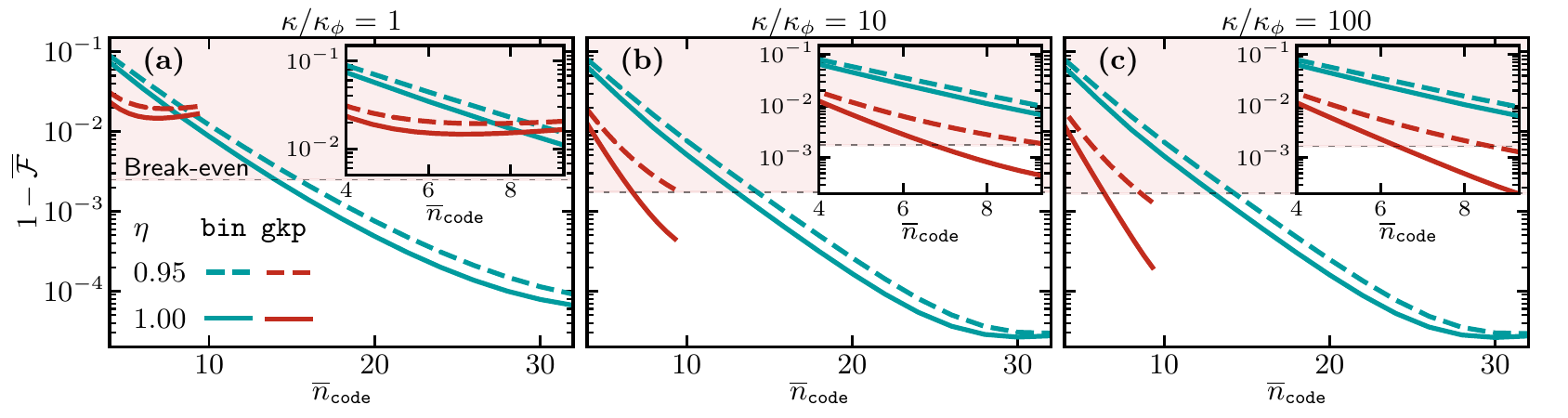}
    \caption{Performance of \qcode{gkp} and \qcode{bin} codes subject to the same initial noise channel described by Eq.~\eqref{eq:master_equation_loss_dephasing} with $\kappa \tau = 5 \cdot 10^{-3}$ and different ratios $\kappa / \kappa_{\phi} \in \lbrace 1, 10, 100 \rbrace$ in panels (a)-(c).
    Solid lines show results for the respective telecorrection circuits (\figref{fig:telecorrection} and \figref{fig:gkp_telecorrection}) with unit efficiency measurements, while dashed lines show the results obtained for a measurement with efficiency $\eta = 0.95$.
    We consider RSB codes of of order $N=4$ and $M=1$ for the data and ancilla qubits in \figref{fig:telecorrection}, respectively, and $\overline{n}_{\qcode{data}} = \overline{n}_{\qcode{anci}}$.
    The logical $\qgate{X}$ measurements are both realized as \qmeas{ahd} measurements.
    For the $\qcode{gkp}$ code we set $\delta_{\mathrm{code}} = \delta_{\mathrm{anci}}$. 
    We are limited to $\overline{n}_{\qcode{gkp}} < 10$ ($\Delta_{\qcode{gkp}}^{\mathrm{(dB)}} \approx 10\, \mathrm{dB}$) for the Fock space simulation due to numerical constraints.
    Insets show a zoom into this region.
    Error correction is beneficial whenever the infidelity falls below the dashed line that indicates the break-even point.
    The results for \qcode{bin} codes are defined only for discrete values of $\overline{n}_{\qcode{code}}$ and we shown continuous lines only for improved readability.}
    \label{fig:ineff_rsb_gkp_comp}
\end{figure*}

We limit our analysis to a single value $\kappa \tau = 5 \cdot 10^{-3}$, but consider different ratios $\kappa / \kappa_{\phi} \in \lbrace 1, 10, 100 \rbrace$ because we found GKP codes to be quite sensitive to photon number dephasing.
We chose $\kappa \tau = 5 \cdot 10^{-3}$ in order to be able to resolve some effects of the noise channel for the accessible parameter range $\delta \gtrapprox 0.225$.
%
The results are shown in \figref{fig:ineff_rsb_gkp_comp}.

The results for \qcode{gkp} codes are computed choosing $\delta_{\qcode{data}} = \delta_{\qcode{anci}}$ and both measurements in \figref{fig:gkp_telecorrection} correspond to homodyne measurements.
While we are restricted to $\Delta_{\qcode{gkp}}^{(\mathrm{dB})} \approx 10\, \mathrm{dB}$ due to computational constraints, we still reach the up-to-date best fault-tolerance threshold of $\Delta_{\qcode{gkp}}^{(\mathrm{dB})} \approx 9.9\, \mathrm{dB}$ computed recently in Ref.~\cite{noh_low-overhead_2022}.
We choose an order-$N=4$ \qcode{bin} code as a representative of the class of RSB codes for the top rail in \figref{fig:telecorrection}.
We choose for the ancilla system an $M=1$ \qcode{cat} code as in \secref{sec:num_res_rsb}, but now with the same  photon number as the data qubit, that is, $\overline{n}_{\qcode{data}} = \overline{n}_{\qcode{anci}}$.
Additionally, we assume that both measurements are realized as adaptive homodyne measurements for the RSB code.
We show results for unit efficiency detectors as well as detectors with efficiency $\eta = 0.95$ and adjust the decision boundaries for the \qcode{gkp} code as described in \secref{sssec:comp_ana_num}.
We deviate here from the procedure of \secref{sec:num_res_rsb} in the above described way in order to create matching conditions for \qcode{gkp} and \qcode{cat} codes.

From \sfigref{fig:ineff_rsb_gkp_comp}{a} we can make the important observation that \qcode{gkp} codes show worse performance in comparison to \qcode{bin} codes if the photon dephasing noise is relatively strong (here $\kappa_{\phi} \tau = 5 \cdot 10^{-3}$).
Thus, even though \qcode{gkp} codes outperform \qcode{cat} codes under a pure loss channel~\cite{albert_performance_2018}, this advantage diminishes in the presence of photon number dephasing.
In \sfigref{fig:ineff_rsb_gkp_comp}{b} and \sfigref{fig:ineff_rsb_gkp_comp}{c} we consider the more optimistic cases with $\kappa / \kappa_{\phi} = 10$ and $\kappa / \kappa_{\phi} = 100$ corresponding to $\kappa_{\phi} \tau = 5 \cdot 10^{-4}$ and $\kappa_{\phi} \tau = 5 \cdot 10^{-5}$, respectively.
In these cases, \qcode{gkp} codes always outperform \qcode{bin} codes when comparing them at the same average photon number $\overline{n}_{\qcode{code}}$.
It is important to note that RSB codes typically achieve their optimal performance only for much larger $\overline{n}_{\qcode{code}}$ in comparison to $\qcode{gkp}$ codes.
Practically, this may not be a limiting factor. For example, \qcode{cat} codes with up to 100 photons have been deterministically prepared in experiments~\cite{vlastakis_deterministically_2013}, while it is still challenging to \timo{prepare} \qcode{gkp} codes with more than $10$ photons~\cite{campagne-ibarcq_quantum_2020, de_neeve_error_2022, eickbusch_fast_2021, kudra_robust_2021}.

Another potential advantage of \qcode{bin} codes over \qcode{gkp} codes becomes apparent when considering measurement of the data qubit with a finite detection efficiency of $\eta = 0.95$ (dashed lines).
On the one hand, an inefficiency of only $5\, \%$ already leads to a significant performance loss for \qcode{gkp} codes of roughly an order of magnitude at $\overline{n}_{\qcode{code}} \approx 10$ in \sfigref{fig:ineff_rsb_gkp_comp}{b} and \sfigref{fig:ineff_rsb_gkp_comp}{c}. 
This performance loss prohibits the \qcode{gkp} code to surpass break-even within our simulation constraints ($\Delta_\qcode{gkp}^{\mathrm{(dB)}}\lesssim 10$ dB).
On the other hand, for \qcode{bin} codes, the loss of performance is negligible close the optimal value $\overline{n}_{\qcode{code}}$.

While in \figref{fig:ineff_rsb_gkp_comp} we have considered only a single value of $\eta \neq 1$, \figref{fig:ineff_fid_bin_N=3} contains additional results for \qcode{bin} codes under a similar noise model, and \figref{fig:rescaled_gkp_fock_twirling} contains additional results for \qcode{gkp} codes in the absence of the noise channel obtained from Eq.~\eqref{eq:master_equation_loss_dephasing}.
From these results we expect the gap between \qcode{gkp} and \qcode{bin} codes to widen further if $\eta$ is decreased more towards realistic values, with $\qcode{gkp}$ codes eventually unable to reach break-even for \emph{any} $\overline{n}_{\qcode{code}}$ as evident from \figref{fig:rescaled_gkp_fock_twirling}.

\section{Discussion and Conclusion\label{sec:discussion_conclusion}}
In this article we have analyzed the performance of teleportation-based error correction circuits for rotation-symmetric bosonic (RSB) codes and Gottesman-Kitaev-Preskill (GKP) codes, 
incorporating finite detection efficiency in the  measurement model.
To this end, we have used exact numerical methods for RSB and GKP codes as well as an analytical model based on the twirling approximation for GKP codes.
Comparing the performance of these bosonic codes within the considered noise model, we have found that RSB codes can outperform GKP codes either in the presence of photon number dephasing or finite measurement efficiencies.

While fault-tolerant thresholds for bosonic encodings reported in the literature seem to reach experimentally feasible squeezing values~\cite{noh_low-overhead_2022}, our study emphasizes the relevance of incorporating realistic noise models for the measurement process as well photon number dephasing into further studies.
%
%
In this sense, our work accompanies previous works that have considered noisy measurement models for GKP error correction schemes, notably Refs.~\cite{vuillot_quantum_2019, noh_fault-tolerant_2020} for gate-based computing and Refs.~\cite{fukui_high-threshold_2019, fukui_all-optical_2021} for measurement-based computing.
However, the results of Refs.~\cite{vuillot_quantum_2019, noh_fault-tolerant_2020} assume that all circuit elements, that is, gates as well as measurements, are comparably noisy.
In particular, the thresholds obtained in Ref.~\cite{noh_fault-tolerant_2020} would require measurements with efficiencies above $99\%$.
Having in mind the currently achievable measurement efficiencies and the results of our work that show that the GKP code is sensitive to finite efficiencies in the measurement, we identify finite measurement efficiencies as a central issue for fault-tolerant quantum computing with GKP codes.
%
For example, for topologically protected measurement-based quantum computing that utilizes post-selection, Ref.~\cite{fukui_high-threshold_2019} numerically established a bound on the minimum efficiency  $\eta \approx 92.2 \,\%$ below which fault-tolerant computation is impossible.
We argue that for the here considered RSB codes~\footnote{We expect the results to hold for any code within the family of number-phase codes.} it is likely that lower measurement efficiencies will be tolerable in fault-tolerant architectures as the phase measurement is not very sensitive to finite measurement efficiencies.
A quantitative analysis of this claim is, however, beyond the scope of this work and is left open for future study.

\timorevised{We note that our results are not specific with respect to the analyzed QEC scheme and can be carried over qualitatively to, for example, concatenated bosonic codes. In fact, our results for different types of measurements as well as the influence of measurement inefficiencies apply qualitatively to any bosonic QEC scheme in which syndrome information is extracted utilizing another bosonic qubit. Examples of such codes are the surface-GKP code (Ref.~\cite{noh_low-overhead_2022} for a teleportation-based version and Ref.~\cite{noh_fault-tolerant_2020} for the gate based version) as well as a foliated version of the surface code~\cite{raussendorf_fault-tolerant_2006, raussendorf_topological_2007, bolt_foliated_2016, brown_universal_2020}  in which data and syndrome qubits are replaced by codes from the class of rotation symmetric bosonic codes.}

We should emphasize that our numerical analysis is still rather idealistic and focuses mostly on the issue of realistic implementations of the phase measurement, restricted to the system that hosts the data qubit.
Additionally, we omit a realistic treatment of state preparation errors of the ancilla mode or errors during the execution of the entangling gates.
However, this noise model can be justified by recognizing that our main goal is to shed light on the role of different currently available measurement schemes when decoding the encoded information.
Furthermore, our model of adaptive homodyne measurements used for the analysis of RSB codes relies on assuming instantaneous adjustment of the local oscillator phase.
Still, previous theoretical~\cite{wiseman_adaptive_1997, berry_effects_2001} as well as experimental~\cite{martin_implementation_2020} results suggest that even with a finite response time, adaptive homodyne detection will be superior to standard heterodyne detection.

We conclude by listing open questions for future research in relation to our work. 
The question to quantify the performance of RSB codes within a fault-tolerant setting, for example based on Bacon-Shor or topological stabilizer codes, remains~\cite{grimsmo_quantum_2020}.
Another interesting avenue is to design numerical methods to simulate RSB codes effectively.
On another note, our scheme further illustrates the need for improved detection schemes in the microwave regime to overcome the poor measurement efficiencies.
One may be able to perform a phase preserving amplification step before releasing the mode to the measurement chain.
Finally, it is also interesting to consider alternative approaches to perform the phase measurement, e.g., by using ancillary qubit systems as they are used in schemes for GKP error correction~\cite{terhal_encoding_2016, campagne-ibarcq_quantum_2020, royer_stabilization_2020}.

\begin{acknowledgments}
We thank Laura García Álvarez,  Oliver Hahn and David Fitzek for useful discussions.
Furthermore, we acknowledge Ingrid Strandberg for useful discussions and comments on the manuscript.
G. F. acknowledges support from the Swedish Research Council (Vetenskapsrådet) Grant QuACVA.
F.Q. and G.F. acknowledge the financial support from the Knut and Alice Wallenberg Foundation through the Wallenberg Centre  for  Quantum  Technology  (WACQT).
T.H. and G.F. acknowledge funding through the Chalmers Excellence Initiative Nano.
A.L.G. is supported by the Australian Research Council, through the Centre of Excellence for Engineered Quantum Systems (EQUS) project number CE170100009 and Discovery Early Career Research Award project number DE190100380.
\end{acknowledgments}

\appendix

\section{Error correction by teleportation\label{sec:knill_ec}}
Error correction by teleportation was put forward by Knill~\cite{knill_quantum_2005, knill_scalable_2005} as a means of increasing the tolerated error probability per gate in fault-tolerant quantum computing, as well as reducing the resource overhead that typical error correction protocols generated.
Here, we briefly recall these ideas as we make considerable use of them in the main text of the article.

\begin{figure}
    \centering
    \begin{quantikz}[column sep=0.625cm, row sep={1.1cm,between origins}]
    \lstick{$\ket{\psi}_1$}   & \qw       & \qw       & \ctrl{1}\gategroup[2,steps=3,style={dashed,rounded corners,fill=Maroon!15, inner xsep=3pt, yshift=3pt, inner ysep=5pt},background]{}  & \gate{H}          & \meter[]{$m_x$} \vcw{2}  \\
    \lstick{$\ket{0}_2$} & \gate{H}\gategroup[2,steps=2,style={dashed,rounded corners,fill=teal!15, inner xsep=2pt, inner ysep=0pt},background]{}   & \ctrl{1}  & \targ{}   & \meter{$m_z$} \vcw{1} \\
    \lstick{$\ket{0}_3$}      & \qw       & \targ{}   & \qw       & \gate{X^{m_z}}    & \gate{Z^{m_x}} & \qw  \rstick{$\ket{\psi}_3$}
    \end{quantikz}
    \caption{Conventional quantum teleportation circuit. In the initial step a Bell pair is created with qubits two and three using the Hadamard and the \qgate{CNOT} gate (blue box). Then a Bell measurement between qubits one and two is performed (red box), teleporting the initial state $\ket{\psi}$ to system three up to Pauli corrections determined by the measurement result $\lbrace m_x, m_z \rbrace$.}
    \label{fig:conv_teleportation_circuit}
\end{figure}
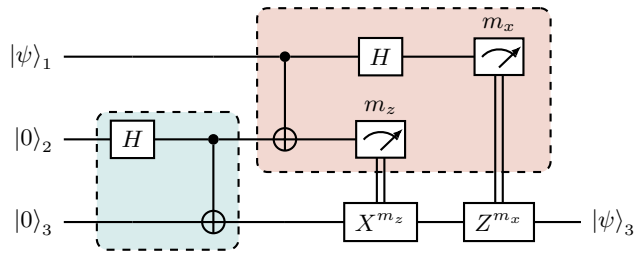

Knill-EC utilizes the conventional quantum teleportation protocol which is depicted in~\figref{fig:conv_teleportation_circuit}.
In the initial step a Bell state $(\ket{00}_{23} + \ket{11}_{23})/\sqrt{2}$ between systems two and three is created (blue box).
Subsequently, using a \qgate{CNOT} and a Hadamard gate as well as measurements in the computational basis, a Bell measurement on the input state and system two is performed with outcome $\lbrace m_x, m_z \rbrace \in \lbrace 0, 1 \rbrace^2$ (red box).
The Bell measurement teleports the input state $\ket{\psi}_1$ to the third system up to a change of the Pauli frame that is known from the measurement result $\lbrace m_x, m_z \rbrace$.
However, this can lead to an Pauli error if either one of the measurement results is erroneous.

To circumvent the issue of Pauli errors due to unreliable measurements, one can perform the teleportation circuit (\figref{fig:conv_teleportation_circuit}) with encoded qubits, replacing all qubits and operations by logical ones.
If the code belongs to the class of stabilizer codes, performing the Bell-basis measurement is equivalent to performing syndrome measurements of the check matrix~\cite{knill_quantum_2005, knill_scalable_2005}.
From the syndrome $\mathbf{e}$ the change in the Pauli frame can be determined.
In particular, for a code with good error-detecting or \nobreakdash-correcting properties, decoding the syndrome $\mathbf{e}$ yields a more reliable result of the Bell-basis measurement and therefore reduces the error of the teleported state in comparison with the input state.
The underlying principle can be elevated from discrete-variable encodings to continuous-variable encodings discussed in the main text, by replacing blocks that encode $n$ physical qubits into a single logical qubit, with a qubit encoded into the continuous degrees of freedom of a physical system.

\section{Error propagation for RSB codes\label{app:rsb_error_propagation}}
Ref.~\cite{grimsmo_quantum_2020} showed that the Kraus operators of \emph{any} single-mode quantum channel can be expanded in the operator basis 
\begin{align}
    \label{eq:error_operator}
    \hat{E}_k(\theta) = \begin{cases}
    e^{i \theta \hat{n}} \hat{a}^{\lvert k \rvert} &\text{for } k < 0, \\
    (\hat{a}^{\dagger})^{\lvert k \rvert} e^{- i \theta \hat{n}} &\text{for } k \geq 0,
    \end{cases}
\end{align}
with $k \in \mathbb{Z}$ representing a shift in the photon number, either upwards (positive) or downwards (negative).
It is therefore sufficient to know how an error $\hat{E}_k(\theta)$ propagates through the gates $\hat{Z}_N$ and $\qgate{CROT}_{N M}$.
Using that $f(\hat{n}) \hat{a} = \hat{a} f(\hat{n} - \mathbb{1})$ for any analytic function $f(\hat{n})$ we find
\begin{align}
    \label{eq:z_error_prop}
    \hat{Z}_N \hat{E}_k(\theta) = e^{i \pi k / N} \hat{E}_k(\theta) \hat{Z}_N,
\end{align}
that is, the $\hat{Z}_N$ gate does not amplify errors.
Furthermore, denoting with an additional superscript $a / b$ the mode an operator acts on, we find for the $\qgate{CROT}$ gate
\begin{align}
    \label{eq:crot_error_prop}
    \qgate{CROT}_{NM} \hat{E}^{a}_{k}(\theta) = \hat{E}^{a}_k(\theta) \hat{E}_0^{b} \left( \frac{k \pi}{N M} \right) \qgate{CROT}_{NM}.
\end{align}
The rotation error $\hat{E}_0^{b} \left(k \pi /N M  \right)$ on mode $b$ that has originated from a number-shift error on mode $a$ is small if $\abs{k} < N / 2$ because in that case the rotation angle $\abs{k} \pi /N M  < \pi / 2 M$ is small with respect to the rotational distance of $\ket{\pm_M}$.

It is also useful to consider the propagation of number-shift errors through the unitary dynamics that are generated by a Hamiltonian that is quadratic in the number operator $\hat{n}$ of a single mode, e.g., the Kerr Hamiltonian $\hat{H}_{K} = K \hat{n}^2$.
The reason is that the Kerr Hamiltonian $\hat{H}_{\chi}$ allows for implementing the logical phase gate $\qgate{S} = \mathrm{diag}(1, i)$ without the need for the teleportation gadget in Eq.~\eqref{eq:gate_teleportation_circuit}~\cite{grimsmo_quantum_2020}.
Commuting $\hat{E}_k(\theta)$ through the unitary generated by this Hamiltonian introduces an additional rotation error~\cite{grimsmo_quantum_2020, albert_performance_2018}
\begin{align}
    \label{eq:kerr_error_prop}
    e^{i \hat{H}_{K}} \hat{E}_k(\theta) \begin{cases}
    e^{-i K k^2} \hat{E}_{k} ( \theta + K k) e^{i \hat{H}_{K}} \quad &\text{for }k < 0, \\
    e^{+i K k^2} \hat{E}_{k} ( \theta + K k ) e^{i \hat{H}_{K}} \quad  &\text{for }k \geq 0.
    \end{cases}
\end{align}
We touch upon the issue of undesired unitary Kerr evolution in \secref{sec:noise_model} in more detail when we discuss the analyzed noise model.

\section{$H_{mn}^{\qmeas{ahd}}$ for adaptive homodyne detection \label{app:h_mat_ahd}}
Here we present for completeness the equation that determines the matrix elements of the $H^{\qmeas{ahd}}$ matrix for adaptive homodyne measurements denoted as the mark II scheme in Ref.~\cite{wiseman_adaptive_1998}.
We do not attempt to reproduce the derivation, only giving the result, namely
\begin{align}
    \label{eq:h_mat_ahd}
    H_{mn}^{\qmeas{ahd}} =  \sum_{p=0}^{\left \lfloor{m/2} \right \rfloor} \sum_{q=0}^{\left \lfloor{n/2} \right \rfloor}
     \gamma_{m, p} \gamma_{n, q} C^{(n, m)}_{p, q},
\end{align}
where $\left \lfloor{m/2} \right \rfloor$ denotes the integer part of $m/2$ and the coefficient $\gamma_{m, p}$ is given by
\begin{align}
    \gamma_{m, p} = \frac{\sqrt{m!}}{2^p (m - 2p)!\, p!}.
\end{align}
$C^{(n, m)}_{p, q}$ can only be expressed in terms of a MacLaurin series expansion~\cite{wiseman_adaptive_1998}.
It then takes the form
\begin{align}
    \label{eq:maclaurin_expansion}
    C^{(n, m)}_{p, q} = \sum_{\ell = 0}^{\infty} \sum_{\ell^{\prime} = 0}^{\infty} \mqty(\frac{n - m}{2} \\ \ell) \mqty(\frac{m - n}{2} \\ \ell^{\prime}) M_{p + \ell, q + \ell^{\prime}},
\end{align}
with 
\begin{align}
    \mqty(\alpha \\ n) = \prod_{k = 1}^{n} \frac{\alpha - k + 1}{k},
\end{align}
the generalized binomial coefficient and the moments $M^{n, m}$ are recursively defined from 
\begin{align}
    M_{n, m} = \frac{n M_{n-1, m} + m M_{n, m-1}}{2 (n-m)^2 + n + m},
\end{align}
with initial values
\begin{align}
    M_{n, 0} = M_{0, n} = \frac{1}{(2n + 1)(2n -1)\dots1} = \frac{1}{(2n + 1)!!}.
\end{align}
In practice, we truncate the series expansion for Eq.~\eqref{eq:maclaurin_expansion} as it converges sufficiently fast.
For large $n, m$ it usually suffices to truncate the sums at $\ell_{\max} = \ell_{\max}^{\prime} \approx \max(n, m)$.
%
%

\section{Equivalence of entanglement fidelity and logical success probability \label{app:entanglement_fid}}
We now show the equivalence of entanglement fidelity and logical error probability for the logical channel arising from the error-correction by teleportation circuit for GKP codes.
Here, we restrict ourselves to the noise model considered in \secref{ssec:gkp_noise_correction}, but the extension to arbitrary noise channels should be straightforward if the noise is agnostic of the explicit code state.

Recall that the entanglement fidelity $\entFid$ of the error-correction circuit (\figref{fig:gkp_telecorrection}) is given by
\begin{align}
    \label{eq-app:entanglement_fid}
    F = \frac{1}{4} \sum_{\vec{a}} \sum_{\vec{x}} P\left( \vec{x} \vert \vec{a} \right) \delta_{\qgate{R}_{\vec{x}} \vert \qgate{P}_{\vec{a}}},
\end{align}
with $P\left( \vec{x} \vert \vec{a} \right) = \Tr[\hat{M}_{x_1} \otimes \hat{M}_{x_2} \mathcal{N}\left(\ketbra{\vec{a}}\right)]$ the conditional probability that a basis state $\ket{\vec{a}}= \ket{a_1} \otimes \ket{a_2}$ subject to the noise $\mathcal{N}(\boldsymbol{\cdot})$ is assigned the measurement outcome $\vec{x} = (x_1, x_2)^{T}$.
The Kronecker delta $\delta_{\qgate{R}_{\vec{x}} \vert \qgate{P}_{\vec{a}}}$ is one whenever the measurement result yields the correct Pauli recovery operation $\qgate{R}_{\vec{x}}$ and zero otherwise.

In the case of closest-integer decoding, the measurement operators $\hat{M}_x$ can be written as
\begin{align}
    \label{eq-app:povm_elements_gkp}
    \hat{M}_{x} = \sum_{n \in \mathbb{Z}} \int_{(2n + x - 1/2) \sqrt{\pi}}^{(2n + x + 1/2) \sqrt{\pi}} \ketbra{p} \dd{p} \equiv \int_{S(x)} \ketbra{p} \dd{p},
\end{align}
where the shorthand $S(x)$ for the integration region denotes the line segments that are associated to the outcome $x$.

The twirling approximation converts coherent Gaussian shifts into incoherent Gaussian shifts so that the logical state takes the form~\cite{noh_fault-tolerant_2020}
\begin{align}
    \label{eq-app:twirling_approximation}
    \hat{\psi}_{\delta} &\propto \mathcal{N}_{\sigma_{\qcode{gkp}}}(\ketbra{\psi}_{\qcode{gkp}}) \\
    &= \int e^{- \frac{s^2 + t^2}{2 \sigma^2_{\qcode{gkp}}}} e^{i s \hat{x}} e^{-i t \hat{p}}  \ketbra{\psi}_{\qcode{gkp}} e^{i t \hat{p}} e^{-i s \hat{x}} \frac{ \dd{s} \dd{t}}{2 \pi \sigma_{\qcode{gkp}}^2},    
\end{align}
where we used Eq.~\eqref{eq:displacement_channel} and $\alpha = \frac{1}{\sqrt{2}}(t + \mathrm{i} s)$.
Intuitively, the twirling approximation results in a grid state where every peak is a physical state with the inconvenience that there are still an infinite number of peaks so that $\hat{\psi_{\delta}}$ still cannot be normalized.
To avoid this issue, let us consider for now a superposition of finitely many states and take the appropriate limit to infinity at the end, that is, we express the logical dual-basis code states as
\begin{align}
    \label{eq-app:finite_superposition_gkp}
    \ket{\pm}_{\qcode{gkp}} =  \lim_{N_{\max} \to \infty} \sum_{n = - N_{\max}}^{N_{\max}} \frac{\ket{p_n^{\pm} = [2 n + (1 \pm 1)/2] \sqrt{\pi}}_p }{\sqrt{2 N_{\max}}},
\end{align}
where $\ket{p}_p$ denotes a momentum eigenstate.

Having established the necessary definitions, we are now in the position to show that within the twirling approximation the entanglement fidelity $\entFid$ and the logical success probability $P_{\mathrm{succ}}(\sigma) = 1 - P_{\mathrm{err}}(\sigma)$ coincide.
First, note that $P(\vec{x}\vert\vec{a})$ factorizes in the case of independent noise, i.e.,
\begin{align}
    \label{eq-app:P_factorization}
    P(\vec{x}\vert\vec{a}) = \Tr_1[M_{x_1} \mathcal{N}(\ketbra{a_1})] \Tr_2[M_{x_2} \mathcal{N}(\ketbra{a_2})].
\end{align}
Performing the integrals one derives the expressions for $\Tr_i[M_{x_i} \mathcal{N}_{\sigma_i}(\ketbra{a_i})]$ by inserting the definitions \eqref{eq-app:povm_elements_gkp}-\eqref{eq-app:finite_superposition_gkp}, e.g.,
\begin{widetext}
\begin{align} 
\Tr \Big[ \hat{M}_{+} & \mathcal{N}_{\sigma_{\qcode{gkp}}}(\ketbra{+}) \Big] = \int_{S(+)}  \bra{p} \mathcal{N}_{\sigma_{\qcode{gkp}}}(\ketbra{+}_{\qcode{gkp}}) \ket{p} \dd{p} \\
&= \lim_{N_{\max} \to \infty}  \frac{1}{2 N_{\max}} \sum_{n, m=N_{\max}}^{N_{\max}} \int_{S(+)} \dd{p} \int_{-\infty}^{\infty} \dd{s} \int_{-\infty}^{\infty} \dd{t} \bra{p} e^{i s \hat{x}} e^{-it\hat{p}} \ketbra{p_n^{+}}{p_m^{+}} e^{i t \hat{p}} e^{-is \hat{x}} \ket{p} \frac{e^{- \frac{s^2 + t^2}{2 \sigma^2_{\qcode{gkp}}}}}{2 \pi \sigma_{\qcode{gkp}}^2} \\
&= \lim_{N_{\max} \to \infty}  \frac{1}{2 N_{\max}} \sum_{n, m=N_{\max}}^{N_{\max}} \int_{S(+)} \dd{p} \int_{-\infty}^{\infty}  \frac{\dd{s}}{\sqrt{2 \pi} \sigma_{\qcode{gkp}}} e^{-s^2 / 2 \sigma_{\qcode{gkp}}^2 - (p_n^{+} - p_m^{+})^2 \sigma_{\qcode{gkp}}^2 / 2} \braket{p}{p_n^{+} + s} \braket{p_m^{+} + s}{p} \\
 &= \lim_{N_{\max} \to \infty}  \frac{1}{2 N_{\max}} \sum_{n=N_{\max}}^{N_{\max}} \int_{S(+)} \frac{\dd{p}}{\sqrt{2 \pi} \sigma_{\qcode{gkp}}} e^{-(p_n^{+} - p)^2 /2 \sigma_{\qcode{gkp}}^2} \\
 &= \frac{1}{\sqrt{2 \pi} \sigma_{\qcode{gkp}}} \int_{S(+)} e^{- p^2 /2 \sigma_{ \qcode{gkp}}^2} \dd{p},
\end{align}

where in the last step we used the shift invariance of the integral to rewrite the $2 N_{\max}$ different integrals as $2 N_{\max}$ identical integrals.
Thus, the result becomes independent of the number of peaks considered in Eq.~\eqref{eq-app:finite_superposition_gkp}.
We then find the entanglement fidelity
\begin{align}
    \entFid 
    &= \frac{1}{4} \sum_{\vec{a}} \sum_{\vec{x}} \Tr_1[M_{x_1} \mathcal{N}_{\sigma_{\mathrm{data}}}(\ketbra{a_1})] \Tr_2[M_{x_2} \mathcal{N}_{\sigma_{\mathrm{anci}}}(\ketbra{a_2})] \delta_{\qgate{R}_{\vec{x}} \vert \qgate{P}_{\vec{a}}} \\
    &= \frac{1}{4} \sum_{s_{\mathrm{data}}, s_{\mathrm{anci}} \in \lbrace +, - \rbrace} \Tr_1[M_{s_{\mathrm{data}}} \mathcal{N}_{\sigma_{\mathrm{data}}}(\ketbra{{s_{\mathrm{data}}}})] \Tr_2[M_{s_{\mathrm{anci}}} \mathcal{N}_{\sigma_{\mathrm{anci}}}(\ketbra{{s_{\mathrm{anci}}}})] 
    \\
    &= \frac{1}{2 \pi \sigma_{\mathrm{data}} \sigma_{\mathrm{anci}}} \left( \int_{S(+)} e^{- \frac{p_1^2}{2 \sigma_{\mathrm{data}}^2}} \dd{p_1} \right) \left( \int_{S(+)} e^{- \frac{p_2^2}{2 \sigma_{\mathrm{anci}}^2 }} \dd{p_2} \right) \\
    &= p_{\mathrm{succ}}(\sigma_{\mathrm{data}}) p_{\mathrm{succ}}(\sigma_{\mathrm{anci}}),
\end{align}
where we used that the Kronecker delta removes one of the sums as there exists only one correct recovery operation for each initial state.
This works because the noise structure assures that the probability of choosing the correct recovery operation is always at least as probable as choosing another recovery operation.
Thus, the entanglement fidelity is equivalent to the logical success probability when the logical states are described within the twirling approximation and any other noise source can be transformed into an random displacement channel $\mathcal{N}_{\sigma}$ with variance $\sigma^2$.
\end{widetext}

\section{Details on the numerical implementation\label{app:numerical_details}}
Our overall numerical implementation builds upon several software packages for \texttt{python}~\cite{johansson_qutip_2013, harris_array_2020, virtanen_scipy_2020, johansson_mpmath_2013, hunter_matplotlib_2007} as well as the \texttt{julia}~\cite{bezanson_julia_2017} programming language.
The code for RSB codes is build upon Ref.~\cite{grimsmo_quantum_2020} which can be found on GitHub~\cite{grimsmo_minimal_2020}.

\subsection{GKP simulation}
Numerically we create regularized GKP states $\ket{\mu_{\delta}}_{\qcode{gkp}} \propto \exp[-\delta^2 (\hat{n} + 1/2)] \ket{\mu}_{\qcode{gkp}}$ through the equivalence relation to a weighted sum of squeezed displaced states given in Ref.~\cite{matsuura_equivalence_2020} which reads
\begin{align}
    \label{eq-app:gkp_finite_energy}
    \begin{split}
    \ket{\mu_{\delta}} \propto \sum_{s \in \mathbb{Z}} &\mathrm{e}^{- \frac{\pi}{2} (2 s + \mu)^2 \tanh{\delta^2}} \hat{D}((2 s + \mu) \sqrt{\pi}) \\
    &\times \hat{S}(- \ln{\sqrt{\sinh{\delta^2} \cosh{\delta^2}}}) \ket{0},
    \end{split}
\end{align}
where $\hat{D}(\alpha) = \exp(\alpha \hat{a}^{\dagger} - \alpha^{*} \hat{a})$ and $\hat{S}(r) = \exp[\frac{r}{2} (\hat{a}^{\dagger 2} - \hat{a}^2)]$ are the displacement and squeezing operators, respectively.

For a large truncation in Fock space the \qgate{CZ} for GKP codes becomes unwieldy to compute through the direct matrix exponential because the resulting operator is not sparse.
It is useful to note the following decomposition of the \qgate{CZ} gate~\cite{kalajdzievski_exact_2021},
\begin{align}
    \qgate{CZ}(s) = BS(\theta) \left[ \hat{S}_1(r) \otimes \hat{S}_2(r) \right] BS(\theta - \pi / 2),
\end{align}
with $s = 2 \sinh r$, $\cos \theta = (1 + \mathrm{e}^{2 r})^{-1/2}$, and $BS(\theta) = \exp[i \theta \left(\hat{a}^{\dagger}_1 \hat{a}_2 + \hat{a}_2^{\dagger} \hat{a}_1 \right)]$ is a beamsplitter between the two modes.
This representation is useful because the Fock space representation of the beamplitter can be computed efficiently with methods for sparse matrices.

Finally, the measurement basis for the closest-integer decoding is constructed by expressing the projector $\ketbra{x}$ in the Fock basis as
\begin{align}
    \label{eq-app:x_proj_fockbasis}
    \ketbra{x} = \sum_{n, m = 0}^{\infty} \frac{e^{-x^2}}{\sqrt{2^{n+m} \pi\, n!\,m! }} H_n(x) H_m(x) \ketbra{n}{m},
\end{align}
with $H_n(x)$ the $n$-th Hermite polynomial.
Using numerical integration we construct the operators $\hat{M}_{x}$ [Eq.~\eqref{eq-app:povm_elements_gkp}] successively from the Fock basis elements $\ketbra{n}{m}$ until a desired accuracy is reached.

\end{document}